\documentclass[11pt]{article}
\RequirePackage[OT1]{fontenc}
\usepackage{amsthm,amsmath,amssymb,amsfonts,bm,enumerate,xcolor,graphicx,natbib,color,url}
\RequirePackage[colorlinks,citecolor=blue,urlcolor=blue]{hyperref}
\usepackage{hyperref}
\usepackage{algpseudocode}
\usepackage{subcaption} 
\usepackage{pifont}
\usepackage{textcmds}
\usepackage{booktabs}
\usepackage{enumitem}
\usepackage{multirow}
\usepackage{booktabs}
\usepackage{adjustbox}
\usepackage{afterpage}
\usepackage{longtable}
\usepackage[usestackEOL]{stackengine}
\usepackage[toc,page]{appendix} 
\usepackage{algorithm,algpseudocode}
\usepackage{lipsum}
\usepackage{setspace}

\usepackage{setspace}
\usepackage{multirow}
\usepackage{etoolbox}
\makeatletter
\makeatletter
\g@addto@macro\bfseries{\boldmath}
\makeatother

\usepackage{pgfplots}
\usepackage{tikz}
\usetikzlibrary{shapes, arrows}

\tikzstyle{connector} = [draw, -latex']

\DeclareMathOperator*{\argmin}{arg\,min}
\DeclareMathOperator*{\argmax}{arg\,max}

\usepackage{ulem} 
\def\AH#1{{\textcolor{red}{#1}}}

\def\AH#1{\footnote{{\textcolor{red}{Arnab\AH{}: #1}}}}
\def\AH#1{{\textcolor{red}{(Arnab: #1)}}}

\def\frac#1#2{{\textstyle{#1\over#2}}}

\def\argmin{{\rm argmin}}

\DeclareSymbolFont{AMSb}{U}{msb}{m}{n}
\DeclareMathSymbol{\Natural}{\mathbin}{AMSb}{"4E}
\DeclareMathSymbol{\Integer}{\mathbin}{AMSb}{"5A}
\DeclareMathSymbol{\Real}{\mathbin}{AMSb}{"52}
\DeclareMathSymbol{\Rational}{\mathbin}{AMSb}{"51}
\DeclareMathSymbol{\Imaginary}{\mathbin}{AMSb}{"49}
\DeclareMathSymbol{\Complex}{\mathbin}{AMSb}{"43} 
\DeclareMathSymbol{\Disk}{\mathbin}{AMSb}{"44} 
\def\bi{\begin{itemize}}
\def\ei{\end{itemize}}
\def\bd{\begin{description}}
\def\ed{\end{description}}
\def\ben{\begin{enumerate}}
\def\een{\end{enumerate}}
\def\bv{\begin{verbatim}}
\def\ev{\end{verbatim}}

\def\hat#1{{\widehat{#1}}}

\newcommand{\indep}{\perp\!\!\!\perp}

\def\2to{{\ {\buildrel 2\over \longrightarrow}\ }}

\def\I1ton{{$I_1,\ldots,I_n$}}
\def\X1ton{{$X_1,\ldots,X_n$}}
\def\Y1ton{{$Y_1,\ldots,Y_n$}}
\def\Z1ton{{$Z_1,\ldots,Z_n$}}
\def\R1ton{{$R_1,\ldots,R_n$}}
\def\e1ton{{$e_1,\ldots,e_n$}}
\def\t1ton{{$t_1,\ldots,t_n$}}
\def\x1ton{{$x_1,\ldots,x_n$}}
\def\y1ton{{$y_1,\ldots,y_n$}}
\def\z1ton{{$z_1,\ldots,z_n$}}

\usepackage[english]{babel}
\usepackage{amsfonts}
\usepackage{amssymb}
\usepackage{graphicx}
\usepackage{enumerate}

\newcommand{\blind}{1}

\begin{document}
\thispagestyle{empty}
\baselineskip=28pt
\vskip 2mm


 \begin{center} 
 {\Large{\bf Spatial wildfire risk modeling using mixtures of tree-based multivariate Pareto distributions}}
\end{center}

\baselineskip=12pt

\vskip 1mm

 \if1\blind
 {
 \begin{center}
 \large
 Daniela Cisneros$^1$, Arnab Hazra$^2$, and Rapha\"el Huser$^1$\\ 
 \end{center}
 \footnotetext[1]{
 \baselineskip=10pt Statistics Program, Computer, Electrical and Mathematical Sciences and Engineering (CEMSE) Division, King Abdullah University of Science and Technology (KAUST), Thuwal 23955-6900, Saudi Arabia. E-mails: daniela.cisnerosarce@kaust.edu.sa; raphael.huser@kaust.edu.sa}
 \footnotetext[2]{
 \baselineskip=10pt Department of Mathematics and Statistics, Indian Institute of Technology Kanpur, Kanpur 208016, India. E-mail: ahazra@iitk.ac.in}
 } \fi

 \baselineskip=16pt
 \vskip 2mm
 \centerline{\today}
 \vskip 2mm


\begin{doublespace}
\noindent    {\large{\bf Abstract.}} 
Wildfires pose a severe threat to the ecosystem and economy, and risk assessment is typically based on fire danger indices such as the McArthur Forest Fire Danger Index (FFDI) used in Australia. Studying the joint tail dependence structure of high-resolution spatial FFDI data is thus crucial for estimating current and future extreme wildfire risk. However, existing likelihood-based inference approaches are computationally prohibitive in high dimensions due to the need to censor observations in the bulk of the distribution. To address this, we construct models for spatial FFDI extremes by leveraging the sparse conditional independence structure of Hüsler–Reiss-type generalized Pareto processes defined on trees. These models allow for a simplified likelihood function that is computationally efficient. Our framework involves a mixture of tree-based multivariate Pareto distributions with randomly generated tree structures, resulting in a flexible model that can capture nonstationary spatial dependence structures. We fit the model to summer FFDI data from different spatial clusters in Mainland Australia and 14 decadal windows between 1999--2022 to study local spatiotemporal variability with respect to the magnitude and extent of extreme wildfires. Our results demonstrate that our proposed method fits the margins and spatial tail dependence structure adequately, and is helpful to provide extreme wildfire risk measures.


\vspace{10pt}

\end{doublespace}

\newpage
\baselineskip=25pt

\vspace{-2mm}

\section{Introduction}
\label{sec:Intro}
Wildfires pose severe threats to human lives, the ecosystem, and properties, resulting in major economic costs \citep{beyers2005wildland, thomas2017costs}. Smoke from wildfires can also lead to poor air quality and high-temperature events \citep{sandberg2003wildland, jhariya2014effects}. With global warming and other triggering factors, wildfires are becoming more frequent, particularly in countries like Australia and the United States \citep{westerling2016increasing, stephenson2013estimating, marlon2012long}. The Black Summer bushfires in Australia in 2019--2020 were particularly devastating, scorching millions of hectares and causing significant loss of wildlife \citep{godfree2021implications}.


To assess wildfire risk over space and time, a variety of fire danger indices have been proposed, such as the McArthur Forest Fire Danger Index (FFDI) used in Australia. These fire indices integrate meteorological and fuel information into a single measure, which is then used by various governmental agencies (e.g., the Australian Bureau of Meteorology) for issuing forecasts and warnings and estimating fire suppression difficulty \citep{van1987development, sirca2006ichnusa, fiorucci2008development}. A number of studies examine the reliability \citep{andrews2003evaluation} and sensitivity \citep{dowdy2010index} of fire danger indices using high-resolution datasets, while others attempt to interpolate them spatially \citep{sanabria2013spatial}, through various statistical machine learning algorithms, such as logistic regression, random forests, and inverse-distance weighting.
However, these classical methods ignore the fact that the most devastating and impactful wildfires often lie in the tail of the distribution. Therefore, for accurate risk assessment and effective wildfire management, it is crucial to develop flexible yet resilient statistical approaches that are specifically designed 
to estimate the risk of extreme (rather than medium) wildfires, under both current and future conditions.
While most studies neglect the tail behavior, a recent exception is the paper by \citet{hazra2018semiparametric}, in which a flexible semiparametric Bayesian model is developed to analyze both the bulk and tail properties of the Fosberg Fire Weather Index (FFWI), with the ultimate goal of producing high-resolution fire risk maps in southern California. However, the spatial dependence model developed therein, which is based on a Dirichlet mixture of skew-$t$ processes, is fairly computationally intensive to fit in high dimensions and is also fitted to the entire dataset (extreme and non-extreme wildfires combined), thus favoring model calibration for the more prevalent non-extreme wildfire events. Another related contribution is \citet{cisneros2021combined}, where a multi-stage multivariate spatial model is proposed to jointly analyze wildfire counts and burnt areas over the entire United States. However, their model again does not explicitly rely on extreme-value theory.


In this work, our primary goal is to develop a new statistical methodology that relies on extreme-value theory and that is computationally efficient with large spatial data, in order to help improve the assessment of extreme wildfires and estimate how their marginal distribution and dependence structure varies over both space and time. In our real data application, we focus on estimating the spatial tail dependence structure of FFDI data over Mainland Australia in a spatio-temporal moving window fashion. Our local estimation approach not only allows us to reduce the effective dimensionality of the problem (thus fitting extreme-value models faster), but more importantly, it also allows us to account for the strong spatio-temporal non-stationarity displayed in the spatial tail dependence structure of FFDI data, and to quantify temporal trends and the effect of climate change on the spatially-aggregated risk of extreme wildfires over the past decades. Furthermore, we note that even if the FFDI only serves as a proxy for wildfire risk, our spatial tail dependence analysis also sheds some light on the (local) severity and spatial extent of extreme wildfires.


To tackle this challenge and estimate spatial wildfire risk in each spatial cluster and time window, we develop a spatial model for high-dimensional data that combines ideas from extreme-value theory and graphical models. The theory of graphical models provides a probabilistic framework for studying conditional independence relationships in a wide variety of contexts \citep{lauritzengraphical,koller2009probabilistic}; one benefit of this framework is that it allows one to construct models with a sparse probabilistic structure, thus facilitating both inference and interpretation. Despite the widespread use of graphical models in various fields and applications, they have been mostly limited to the multivariate Gaussian distribution \citep{ren2015asymptotic,epskamp2018gaussian}. However, Gaussian models are unsuitable for studying extreme events in risk analysis due to their thin tails and restrictive dependence structure, which can lead to heavily underestimated probabilities of concurrent extremes \citep{davison2013geostatistics}. Therefore, developing multivariate or spatial graphical models specifically designed for capturing tail dependence in high-dimensional data is crucial. In the framework of max-stable distributions, \citet{gissibl2018max} and \citet{kluppelberg2019bayesian} developed max-linear models on directed acyclic graphs. However, such models have a discrete underlying \emph{angular probability measure}, which is typically unrealistic for spatial data, and \cite{papas2016conditional} showed that density factorization of the most popular max-stable distributions (with continuous angular density) on graphs is not possible, except in trivial cases. 
Another perspective on multivariate extremes, however, is based on threshold exceedances---instead of maxima---, leading to the multivariate Pareto distribution (MPD)---instead of max-stable distributions---occurring in the limit as the threshold grows arbitrarily; see, e.g., \citet{rootzen2006multivariate} and \citet{rootzen2018multivariate}. For MPDs, \citet{engelke2020graphical} introduced new notions of conditional independence and extremal graphical models, which enable data density factorization through a Hammersley--Clifford-type theorem; see also \citet{segers2020one} and \citet{asenova2021inference} for related work on limits of regularly-varying Markov trees. While providing a flexible framework for multivariate extremes, the single tree-based extremal graphical model of \citet{engelke2020graphical} is often unrealistic for spatial data; \citet{engelke2020graphical} also considered block graphs as simple extensions to tree structures, but these remain overly restrictive in realistic spatial settings. In some related work, \citet{yu2016modeling} developed a copula model for station-wise block maxima constructed from an ensemble of trees; however, the linking copulas used in this paper are not motivated by extreme-value theory. Additionally, the analytical expression of the extremal dependence $\chi$-measure \citep{coles1999dependence} is intractable for their proposed model; only a lower bound depending on the underlying tree structure is derived.

In this paper, we model high FFDI threshold exceedances over space using a tree-based H\"usler--Reiss-type generalized Pareto process mixture. For simplicity, we estimate the marginal distribution parameters and the spatial dependence parameters separately in two stages. For marginal estimation (first stage), we borrow information from the neighboring locations to obtain spatially-smooth estimates of the parameters of the location-specific generalized Pareto distributions; using these estimates, we then transform the margins to the unit Pareto scale. For dependence structure estimation (second stage), we build upon \citet{engelke2020graphical} and \citet{yu2016modeling} and fit a tree-based H\"usler--Reiss-type Pareto process model to the transformed data. While a single tree is unable to properly capture the spatial dependence, a mixture of (appropriately constructed) trees has considerably higher flexibility and is, therefore, better suited for modeling spatial data; moreover, this mixture modeling approach keeps at the same time the computational benefit that comes with the use of tree structures. Thus, our approach is not only justified by multivariate extreme-value theory, but it is also amenable to high spatial dimensions. The resulting model enjoys high flexibility and can in particular capture nonstationary spatial extremal dependence structures by adapting the tree mixture weights in a data-driven way. Furthermore, unlike \cite{yu2016modeling}, we can here obtain the analytical expression of the extremal dependence $\chi$-measure. A simulation study showcases the flexibility of our model in capturing the extremal dependence and the upper tail behavior; see the supplementary material of this paper. For our data application, we split Mainland Australia into 25, 50, and 100 spatial clusters and fit our model locally to summer (Nov., Dec., Jan., Feb.) FFDI data available across 23 summer seasons between 1999–2022. To understand the temporal trends in marginal and dependence structures across Australia, we split the 23 seasons into 14 overlapping decadal windows and fit our model separately for each cluster-window combination. Because high spatial aggregates indicate a higher probability of large-scale extreme fire events, we assess wildfire risk across the spatial clusters and study their temporal evolution.

The rest of the paper is organized as follows. In Section~\ref{sec:DataDescr}, an exploratory analysis of the Australian FFDI dataset is presented. Section~\ref{sec:Methodology} describes our proposed mixture model based on tree-based H\"usler--Reiss-type multivariate Pareto distributions. In Section~\ref{sec:computation}, we discuss model fitting. 
In Section~\ref{sec:Application}, we apply the tree-based mixture model to analyze the Australian FFDI dataset and discuss the results. Section~\ref{sec:Conclusions} concludes and points out some future research directions. 

\section{Australian McArthur Forest Fire Danger Index data} %
\label{sec:DataDescr}

The Australian McArthur Forest Fire Danger Index (FFDI) dataset that we analyze in this paper contains daily fire index observations obtained using weather forecasts from historical simulations. The FFDI data are made publicly available by the European Forest Fire Information System (EFFIS, \url{https://effis.jrc.ec.europa.eu/}) of the European Centre for Medium-Range Weather Forecasts (ECMWF) Re-Analysis, 5th generation (ERA5). The FFDI is a standard measure used across Australia for examining the influence of near-surface weather conditions on fire behavior, with the Australian Bureau of Meteorology routinely issuing FFDI forecasts for use by fire management authorities throughout Australia. Mathematically, FFDI is defined as a nonlinear function of temperature (T [$^{\circ}$C]), relative humidity (RH [\%]), and wind speed ($\nu$ [km/hr]) on a given day, as well as a dimensionless number representing fuel availability, called the drought factor (DF; \citealp{griffiths1999improved}); precisely, it is calculated as 
\begin{eqnarray}\label{eq:ffdi_1}
\text{FFDI} = 2\exp\{-0.45+0.987\,\log(\text{DF})-0.0345\,\text{RH}+0.0338\,\text{T}+0.0234\,\nu\}.
\end{eqnarray}
This formula is a computationally efficient rearranged definition of the commonly used formulation due to \cite{noble1980mcarthur}. The Australian Bureau of Meteorology considers an FFDI value in the interval 0--11 to be low-to-moderate, 12--31 to be high, 32--49 to be very high, 50--74 to be severe, 75--99 to be extreme, and above 100 to be catastrophic.

\begin{figure}[p]
    \centering
\includegraphics[width=1\linewidth]{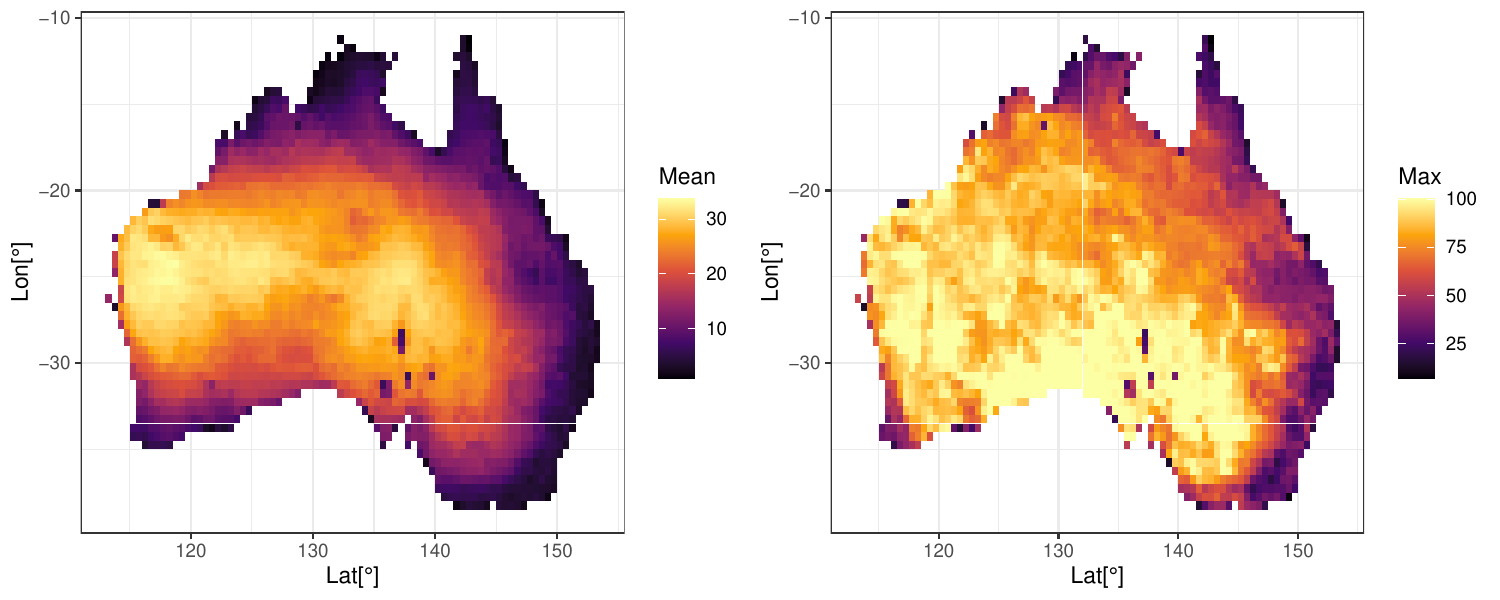}       
    \caption{Grid cell-wise empirical mean (left) and maximum (right) of FFDI data.}
    \label{fig:FFDI_mean_max}
    \vspace{-4mm}
\end{figure}

We limit our study to the Australian mainland that has been seriously affected by wildfires in recent years. We perform some pre-processing steps, which include thinning the grid cells by keeping every second grid cell in each direction to avoid nearly-identical values in neighboring locations (due to wildfires being typically smooth large-scale phenomena), which can cause numerical instabilities. 
The daily FFDI data are then extracted over the years 1999--2022 across Mainland Australia at a spatial resolution of $0.5^{\circ}\times 0.5^{\circ}$, yielding a total of 2749 grid cells across the study domain. We focus on the summer months (Nov., Dec., Jan., and Feb.), which are more prone to wildfires; thus, we analyze data across 23 summer seasons. Figure~\ref{fig:FFDI_mean_max} displays the mean and maximum FFDI values, calculated at each grid cell across the entire time period. While the mean FFDI profile is highest near the mid-western region of Australia, the maximum profile shows higher values near the southern coastal areas; these different spatial patterns indicate that the FFDI data behave differently in the bulk and the tail, suggesting that extreme wildfire risk should be assessed using a specialized tail model fitted to the largest observations only. Interestingly, maximum FFDI values fall in the extreme or catastrophic category for a major part of Mainland Australia, except for coastal regions in the Eastern part of the country, as well as a small region in the South-West. 

\begin{figure}[t!]
    \centering
\includegraphics[width=1\linewidth]{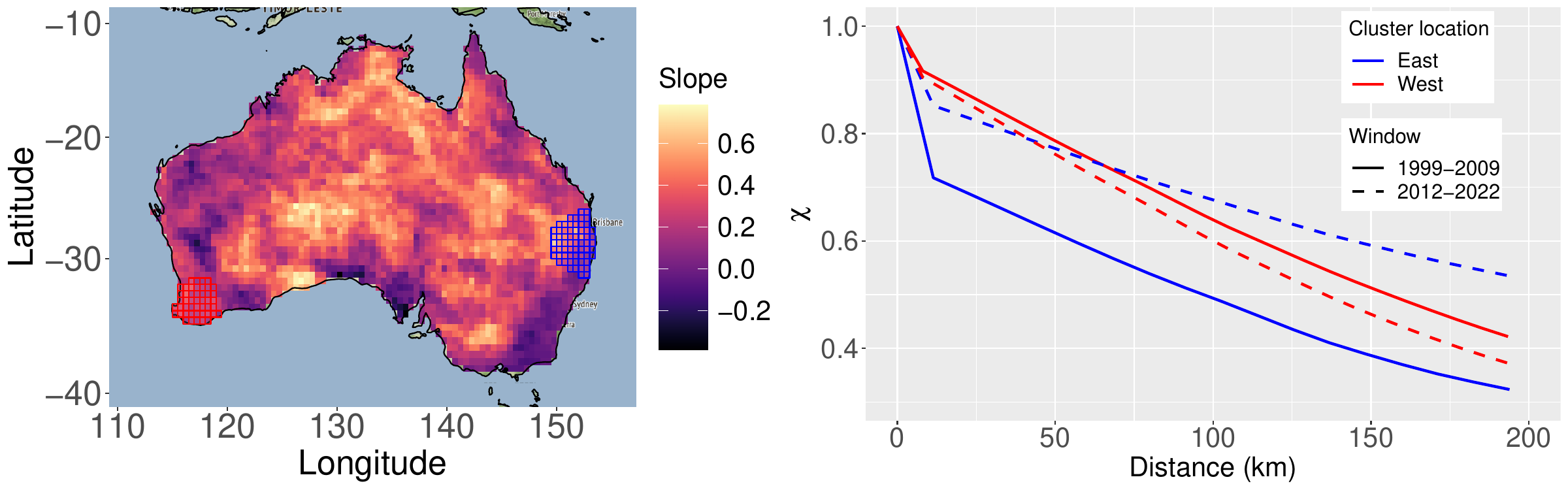}       
    \caption{Grid cell-wise annual rate of change in the $98\%$-empirical quantile (left) and smoothed empirical extremal dependence profiles for two spatial clusters near the western and eastern coasts (highlighted in red and blue in the left panel), for the first and last decadal windows (right).}
    \label{fig:FFDI_change_margin_dep}
    \vspace{-4mm}
\end{figure}

To explore temporal trends in the tails of the FFDI distribution, we compute, for each grid cell and summer season, the $98\%$-empirical quantile and we then fit a simple linear regression model with the season number as a covariate. We present the estimated annual rate of change at each grid cell in the left panel of Figure~\ref{fig:FFDI_change_margin_dep}. While the estimated slope coefficient (i.e., annual rate of change) varies between $-0.39$ and $0.78$, it is positive at 2600 grid cells and negative at 149 grid cells only. Based on grid cell-wise $t$-tests, the slope coefficients are found to be significant at 536 grid cells (i.e., over about $20\%$ of the country), and all of them are significantly positive. We obtain similar results for other high quantiles. This exploratory analysis shows that there has been a significant positive shift in FFDI extremes over the past two decades, and this motivates us to split the 23 summer seasons into 14 overlapping decadal windows and to perform window-specific analyses.

Finally, we also investigate the local spatial extremal dependence structure of FFDI data across Australia and assess the extent of spatial nonstationarity. Mathematically, extremal dependence between two random variables $Y_1 \sim F_1$ and $Y_2 \sim F_2$ can be described by the $\chi$-measure,  $\chi_{12} = \lim_{u \rightarrow 1} \chi_{12}(u)$, where $ \chi_{12}(u) =  \textrm{Pr} \left\{Y_1 > F_1^{-1}(u) \mid Y_2 > F_2^{-1}(u) \right\}$. 
In the spatial context, the $\chi$-measure is often used to discriminate between processes that are asymptotically dependent (with $\chi_{12}>0$ for all pairs of sites) and asymptotically independent (with $\chi_{12}=0$ for all pairs of sites); see \citet{huser2020advances} for an in-depth discussion. Since wildfires are large-scale events, we assume hereafter that the local dependence structure is asymptotically dependent. In case replications of the variables $Y_1$ and $Y_2$ are available, we can easily estimate $\chi_{12}$ by setting a high threshold $u\approx 1$ and calculating the empirical counterpart of $\chi_{12}(u)$. 
Since we have spatial data with temporal replicates, we perform the above-mentioned procedure for every pair of sites $\{\bm s_1,\bm s_2\}$ and then study the pairwise empirical $\chi_{12}$ as a function of the distance $h=\|\bm s_1-\bm s_2\|$ between sites. Here, we split Mainland Australia into 50 different non-overlapping spatial clusters obtained by $k$-means clustering and estimate the local cluster-specific $\chi_{12}$ (assuming local stationarity and isotropy) for each decadal window. The right panel of Figure~\ref{fig:FFDI_change_margin_dep} displays smoothed $\chi_{12}$ function estimates plotted with respect to geodesic distance $h$ (in km) for two clusters---one near the eastern coastal region and the other one near the southwestern coastal region, centered at the coordinates $(151.56^\circ \text{E}, 30.21^\circ \text{S})$ and $(117.67^\circ \text{E}, 33.98^\circ \text{S})$, respectively---for both the first and the last decadal time windows. For the first window (1999--2009), the $\chi_{12}$ values are larger for the cluster near the southwestern coast, while for the last window (2012--2022), the $\chi_{12}$ profiles for the two clusters cross each other. This indicates a nonstationary dependence structure as a whole across Australia and it motivates us to model the data using a local cluster-specific approach. For the eastern cluster, we notice that the empirical $\chi_{12}$ profile is higher for the last decadal window than for the first window. The opposite pattern can be noticed for the other southwestern cluster, though the observed change is much weaker than that for the eastern cluster. Since the $\chi_{12}$ profile increases more rapidly near the eastern coast, where the majority of the largest cities of Australia are located, studying these temporal changes based on a theoretically-justified spatial statistical model is of utmost importance.


\section{Methodology}
\label{sec:Methodology}


Spatial extremes modeling is typically done in two subsequent stages. First, parameters of marginal distributions are estimated by fitting extreme-value theory-justified limit distributions to sample extremes---here, high threshold exceedances. Then, the data are transformed to the same scale using the probability-integral transform. Second, the dependence structure is inferred by fitting a suitable multivariate or spatial extreme-value model. We adopt such a two-stage inference approach to study the behavior of spatial FFDI extremes within each spatial cluster and decadal window. Section~\ref{sub:Stage1} describes marginal modeling, while Section~\ref{sub:Stage2} presents our proposed spatial dependence model based on tree-based MPD mixtures and discusses its properties.

\subsection{Marginal modeling} 
\label{sub:Stage1}

Here we describe the estimation of the marginal distribution at each site based on classical extreme-value theory \citep{davison2015statistics}. Suppose $Z_t(\cdot)$, $t=1,\ldots,T$, denote the $T$ replicates of a spatial stochastic process observed at spatial locations $\bm{s}_i$, $i=1,\ldots,N$. When the process is observed only over a discretized domain, we write $Z_{i,t} \equiv Z_t(\bm{s}_i)$ for simplicity, and we write the set of available observations at site $\bm{s}_i$ as $\mathcal{Z}_i = \{ Z_{i,1}, \ldots, Z_{i,T} \}$. Here the elements of $\mathcal{Z}_i$ form a random sample from some marginal distribution $F_i(\cdot)$ and suppose the Pickands--Balkema--de Haan Theorem \citep{balkema1974residual,pickands1975statistical} holds; that is, we assume that there exists a normalizing function $a(u_i)>0$ such that
\begin{equation}
\label{eq:PdHB.Theor}
\textrm{P}\left(\dfrac{Z_{i,t}-u_i}{a(u_i)}>z \;\;\middle\vert\;\; Z_{i,t}>u_i \right)\rightarrow 1 - H_i(z),\qquad \textrm{as}~~~ u_i\to z_{i\star},
\end{equation}
where $z_{i\star}$ is the upper endpoint of the support of $F_i(\cdot)$, and $H_i(\cdot)$ is a non-degenerate distribution. In that case, the limit $H_i(\cdot)$ must be a generalized Pareto (GP) distribution function \citep{davison1990models} and has the form
\begin{eqnarray}
\label{eq:GP}
H_i(z) = \begin{cases} 1-\left(1+\xi_i z/\sigma_i\right)^{-1/\xi_i}_{+}, &\text{if}~~\xi_i\neq 0, \\ 
1-\exp\left(-z/\sigma_i\right),&\text{if}~~\xi_i=0,
\end{cases}
\end{eqnarray} 
where $(a)_+ = \max\{0, a\}$, and $\sigma_i>0$ and $\xi_i \in \mathbb{R}$ are scale and shape parameters, respectively; we denote this distribution by $\textrm{GP}(\sigma_i, \xi_i)$. The shape parameter $\xi_i$ determines the behavior of the GP distribution: if $\xi_i<0$, $H_i(\cdot)$ has an upper bound at $-\sigma_i/\xi_i$, and if $\xi_i\geq 0$, the upper tail of $H_i(\cdot)$ is unbounded and can be either light ($\xi_i=0$) or heavy ($\xi_i>0$). By replacing the limit in \eqref{eq:PdHB.Theor} with a equality in distribution for some high site-specific threshold $u_i$, and letting the term $a(u_i)$ be absorbed into the GP scale parameter $\sigma_i$, we then obtain the approximation
\begin{eqnarray}
\label{eq:exceed_chap4}
\textrm{P}(Z_{i,t}>z)\approx \zeta_{u_i}\left(1+\xi_i \dfrac{z-u_i}{\sigma_i} \right)^{-1/\xi_i}, \qquad z > u_i,\quad\xi_i\neq0,
\end{eqnarray}
where $\zeta_{u_i}=\textrm{P}(Z_{i,t}>u_i)$, and thus, the level $z_i^{(m)}$ that is exceeded on an average once every $m$ observations is given by
\begin{align}
\label{eq:retlev}
z_i^{(m)}=\begin{cases}
u_i+\dfrac{\sigma_i}{\xi_i}\left\{(m\zeta_{u_i})^{\xi_i}-1\right\},\hspace{.5cm}&\text{if}~~\xi_i \neq 0,\\
u_i+\sigma_i\log\left(m\zeta_{u_i}\right) ,\hspace{.5cm}&\text{if}~~\xi_i=0,
\end{cases}
\end{align}
provided $m$ is sufficiently large to ensure that $z_i^{(m)}>u_i$. The high quantile $z_i^{(m)}$ is commonly called the $m$-observation return level, though its meaning is less clear under temporal non-stationarity.

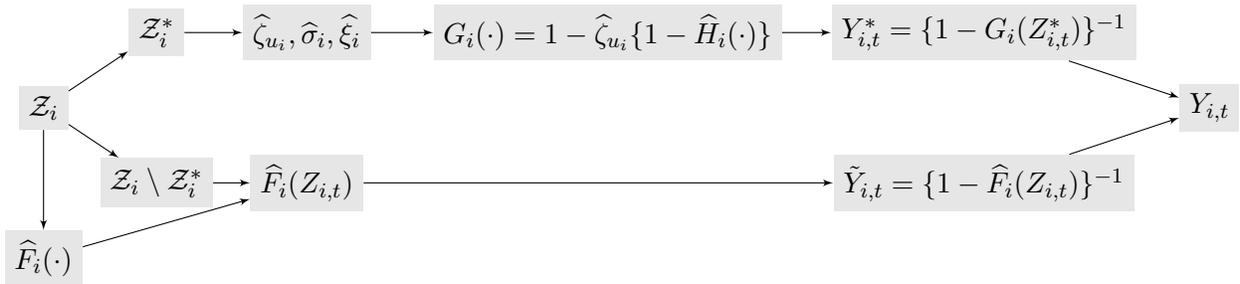
\begin{figure}[t!]
    \centering
    \begin{tikzpicture}
\node [fill=gray!20] at (0,-1) (fulldata) {$\mathcal{Z}_i$};
\node [fill=gray!20] at (1.5,0) (threxceed) {$\mathcal{Z}^*_i$};
\node [fill=gray!20] at (1.5,-2) (belowthr) {$\mathcal{Z}_i \setminus \mathcal{Z}^*_i$};
\node [fill=gray!20] at (3.5,0) (mles) {$\hat{\zeta}_{u_i}, \hat{\sigma}_i, \hat{\xi}_i$};
\node [fill=gray!20] at (7.5,0) (cdfexceed) {$G_i(\cdot) = 1 - \hat{\zeta}_{u_i}\{1 - \hat{H}_i(\cdot)\}$};
\node [fill=gray!20] at (0,-3) (ecdf) {$\hat{F}_i(\cdot)$};
\node[fill=gray!20] at (3.5, -2) (pseudounif) {$\hat{F}_i(Z_{i,t})$};
\node[fill=gray!20] at (12.5, 0) (ystar) {$Y^*_{i,t} = \{1 - G_i(Z^*_{i,t})\}^{-1}$};
\node[fill=gray!20] at (12.5, -2) (ytilde) {$\tilde{Y}_{i,t} = \{1 - \hat{F}_i(Z_{i,t})\}^{-1}$};
\node[fill=gray!20] at (15.5, -1) (unitpareto) {$Y_{i,t}$};
\path [connector] (fulldata) -- (threxceed);
\path [connector] (fulldata) -- (belowthr);
\path [connector] (threxceed) -- (mles);
\path [connector] (fulldata) -- (ecdf);
\path [connector] (ecdf) -- (pseudounif);
\path [connector] (cdfexceed) -- (ystar);
\path [connector] (pseudounif) -- (ytilde);
\path [connector] (mles) -- (cdfexceed);
\path [connector] (belowthr) -- (pseudounif);
\path [connector] (ystar) -- (unitpareto);
\path [connector] (ytilde) -- (unitpareto);
\end{tikzpicture}
    \caption{Flow chart of the marginal estimation procedure, and the transformation of data to the unit Pareto scale.}
    \label{fig:flowchart_stage1}
\end{figure}

To estimate marginal parameters at each site $\bm{s}_i$, we adopt a local inference approach, whereby the GP model is fitted by maximum likelihood by pooling information from a set of sites $\{\bm{s}_j; j\in\mathcal{N}_i\}$ containing $\bm{s}_i$ and its four nearest neighbors. Using data from neighboring sites allows us to increase the effective sample size, thus reducing estimation uncertainty, and to smooth marginal parameters spatially. For each site $\bm{s}_i$, $i=1,\ldots,N$, we fix a high threshold $u_i$ and denote by $\mathcal{Z}^*_j = \{Z^*_{j,1}, \ldots, Z^*_{j,T_j} \} \subset \mathcal{Z}_j$ the $T_j$ values exceeding $u_i$ at site $j\in\mathcal{N}_i$. Under the working assumption of independence across space and time, we then maximize the following likelihood function 
\begin{equation}\label{neighboring_like}
L_i(\sigma_i, \xi_i \mid Z^*_{j,k}, j\in \mathcal{N}_i, k=1,\ldots, T_j) = \prod_{j \in \mathcal{N}_i} \prod_{k=1}^{T_j} \left\{\dfrac{1}{\sigma_i}\left( 1+\xi_i\dfrac{Z^*_{j,k}-u_i}{\sigma_i}\right)\right\}_{+}^{-1/\xi_i-1},
\end{equation}
and we optimize it numerically to obtain the maximum likelihood estimates (MLEs) $\hat{\sigma}_i$ and $\hat{\xi}_i$. Plugging the MLEs into $H_i(\cdot)$ yields the fitted GP distribution $\hat{H}_i(\cdot)$. The probability $\zeta_{u_i}$ can be estimated empirically by $\hat{\zeta}_{u_i} = T_i / T$. We can then approximate the upper tail of the distribution $F_i(\cdot)$ using \eqref{eq:exceed_chap4} as $F_i(z)\approx G_i(z):=1 - \hat{\zeta}_{u_i}\{1 - \hat{H}_i(z)\}$, for $z$ large. For values below the threshold, the empirical distribution function $\hat{F}_i(\cdot)$ can be used instead. Therefore, the full distribution $F_i(z)$ can be estimated by $G_i(z)$, if $z>u_i$, and $\hat{F}_i(z)$, if $z\leq u_i$, and data can then be transformed to the unit Pareto scale via the probability integral transform, $Y_{i,t}=\{1-F_i(Z_{i,t})\}^{-1}$, after replacing $F_i(\cdot)$ with its approximation. The whole marginal estimation procedure and standardization to a common unit Pareto scale are summarized in a flowchart displayed in Figure~\ref{fig:flowchart_stage1}.

\subsection{Extremal dependence modeling} 
\label{sub:Stage2}

\subsubsection{Multivariate Pareto distributions}
\label{subsub:mgpd}
Our spatial dependence model is constructed from mixtures of tree-based multivariate Pareto distributions (MPDs) of H\"usler--Reiss type, so we give a brief background on MPDs, initially introduced by \citet{rootzen2006multivariate} and then further studied by \citet{rootzen2018multivariate}, as well as the specific H\"usler--Reiss model, first introduced by \citet{husler1989maxima} and recently popularized by \citet{engelke2020graphical} in the context of extremal graphical models.


A random vector $\bm{Y} = (Y_{1}, \ldots, Y_{N})'$ with unit Pareto margins is said to be in the max-domain of attraction of the random vector $\bm{W} = (W_1,\ldots, W_N)'$ if, for any $\bm{w} = (w_1, \ldots, w_N)'\in \mathcal{F}=[0, \infty)^N \setminus \{\bm{0}\}$, we have
\begin{equation}\label{exponent}
   \lim_{T\rightarrow \infty} \textrm{P}\left(\max_{t=1, \ldots, T} Y_{1,t} \leq T w_1, \ldots, \max_{t=1, \ldots, T} Y_{N,t} \leq T w_N\right) = \textrm{P}(\bm{W} \leq \bm{w}) =  \exp\{-V(\bm{w})\},
\end{equation}
where $\bm{Y}_t=(Y_{1,t},\ldots,Y_{N,t})'$, $t=1,2,\ldots$, are independent copies of $\bm{Y}$. If this convergence holds, $\bm{W}$ must be max-stable \citep{Davison.etal:2019} with standard Fr\'echet margins, i.e., $\textrm{P}(W_i \leq w) = \exp(-1/w)$, $w \geq 0$, and $V(\bm{w}):=\Lambda([0,\infty]^N\setminus[\bm{0},\bm{w}])$ is known as the exponent function and is constructed from a Radon measure $\Lambda$ defined on the cone $\mathcal{F}$. 
If $\Lambda(\cdot)$ is absolutely continuous with respect to Lebesgue measure on $\mathcal{F}$, its Radon–Nikodym derivative is denoted by $\lambda(\cdot)$ and satisfies $\lambda(\bm{w})=-\dfrac{\partial^N}{\partial w_1 \cdots \partial w_N}V(\bm{w})$. It can be shown that the intensity $\lambda(\cdot)$ has two main properties: (i) it is homogeneous of order $-(N + 1)$, i.e., $\lambda(c\bm{w}) = c^{-(N+1)}\lambda(\bm{w})$ for any $c > 0$ and $\bm{w} \in \mathcal{F}$; and (ii) it has normalized margins in the sense that for any $i \in \{1,\ldots, N\}$, $\int_{\{\bm{w} \in \mathcal{F}: w_i > 1\}} \lambda(\bm{w}){\rm d}\bm{w} = 1$.

Following \cite{resnick2008extreme}, the convergence in (\ref{exponent}) is equivalent to
$\lim_{u\rightarrow \infty} u\{1 - \textrm{P}(\bm{Y} \leq u \bm{w})\} = V(\bm{w}), \bm{w} \in \mathcal{F}$; thus, for $\bm{w} \in \mathcal{F}$, the joint distribution of the multivariate threshold exceedances of $\bm{Y}$, defined through their infinity norm being large, satisfies
\begin{equation}\label{cdf_exceedance}
   \textrm{P}\left(\bm{X} \leq \bm{x}\right) = \lim_{u\rightarrow \infty} \textrm{P}\left(\bm{Y} / u  \leq \bm{x} \mid \lVert \bm{Y} \rVert_{\infty} > u\right) =  [V(\min\{\bm{x}, \bm{1}\}) - V(\bm{x})]/V(\bm{1}),
\end{equation}
where $\min\{\bm{x}, \bm{1}\}$ denotes the vector of component-wise minima between $\bm{x}$ and $\bm{1}$. The limiting random vector $\bm{X}$ is said to follow an MPD \citep{rootzen2006multivariate}. The underlying exponent measure $\Lambda(\cdot)$ of the distribution of $\bm{W}$ implicitly defines the MPD of $\bm{X}$, with support $\mathcal{L} = \{\bm{x} \in \mathcal{F}: \lVert \bm{x} \rVert_{\infty} > 1\}$. Further, we assume that the distribution of $\bm{X}$ allows a positive and continuous density $f_{\bm{X}}(\cdot)$ on $\mathcal{L}$ given by
$f_{\bm{X}}(\bm{x}) = \dfrac{\partial^N}{\partial x_1 \cdots \partial x_N} \textrm{P}\left(\bm{X} \leq \bm{x}\right) = \lambda(\bm{x})/V(\bm{1})$ for $\bm{x} \in \mathcal{L}$; the mixed derivative of $V(\min\{\bm{x}, \bm{1}\})$ in (\ref{cdf_exceedance}) cancels out because it is constant with respect to at least one $x_i$ as $\lVert \bm{x} \rVert_{\infty} > 1$. The intensity $\lambda(\cdot)$ can thus be interpreted as the unnormalized MPD density, while the normalization constant $V(\bm{1})=\Lambda([0,\infty]^N\setminus[\bm{0},\bm{1}])$ ensures that it integrates to one over $\mathcal{L}$. Here $V(\bm{1})$ is called the $N$-variate extremal coefficient and serves as a summary of extremal dependence.

A particularly interesting parametric multivariate extreme-value model is the H\"usler--Reiss distribution, which arises as the only possible limit of renormalized maxima from suitably defined triangular arrays of multivariate Gaussian vectors \citep{husler1989maxima}. Because of this connection with the multivariate Gaussian distribution, the H\"usler--Reiss model or its spatial extension known as the Brown--Resnick process \citep{kabluchko2009stationary} have been found to provide good fits in practical applications. Moreover, there are further fundamental links with the multivariate Gaussian distribution, especially in the context of graphical models \citep{engelke2020graphical}, that make them appealing, and we shall explain in Section~\ref{subsub:single_tree} how to construct tree-based H\"usler--Reiss MPDs. A $N$-dimensional H\"usler--Reiss distribution is parameterized by a variogram matrix $\bm{\Gamma}$ (i.e., symmetric, strictly conditionally negative-definite with non-negative entries $\Gamma_{ij}$ and diagonal entries $\Gamma_{ii} = 0$, $i,j=1,\ldots,N$). For any $k \in \{1, \ldots, N\}$, the corresponding intensity $\lambda(\cdot;\bm \Gamma)$ is
\begin{equation} \label{eq:lambda_hr}
    \lambda(\bm{x};\bm \Gamma)=x_k^2\prod_{i\neq k}x_i^{-1}\phi_{N-1}\left(\tilde{\bm{x}}_{\setminus k};\bm{\Sigma}^{(k)}\right),~ \bm{x} \in \mathcal{L},
\end{equation}
where $\tilde{\bm{x}}=[\log(x_1/x_k)+\Gamma_{1k}/2, \ldots, \log(x_N/x_k)+\Gamma_{Nk}/2]'$, $\tilde{\bm{x}}_{\setminus k}$ denotes the vector $\tilde{\bm{x}}$ with the ${k\mbox{-th}}$ element removed, and $\phi_{N-1}\left(\cdot~;\bm{\Sigma}^{(k)}\right)$ is the $(N-1)$-dimensional centered multivariate normal density with the strictly positive-definite covariance matrix
\begin{equation} \label{lambda_covmat}
    \bm{\Sigma}^{(k)}=\dfrac{1}{2}\left\lbrace\Gamma_{ik}+\Gamma_{jk}-\Gamma_{ij}\right\rbrace_{i,j\neq k}\in\mathbb{R}^{(N-1)\times(N-1)}.
\end{equation}
Note that, although the right-hand side of the density in (\ref{eq:lambda_hr}) appears to depend on the choice of $k$, it is actually independent of $k$. The strength of dependence between the ${i\mbox{-th}}$ and ${j\mbox{-th}}$ component of a vector distributed according to the H\"usler--Reiss model is thus controlled by $\Gamma_{ij}$; here $\Gamma_{ij} = 0$ leads to complete dependence between the components while the dependence weakens as $\Gamma_{ij}\to\infty$.

\subsubsection{Tree-based multivariate Pareto distributions}
\label{subsub:single_tree}

We now briefly recall the approach of \cite{engelke2020graphical} for creating graphical models for multivariate extremes, and tree-based H\"usler--Reiss MPD models.

An undirected graphical model induces a multivariate density function $f(\bm x)$ that factorizes according to a graph $\mathcal{G}$, defined by a collection of nodes $\mathcal{V}$ and edges $\mathcal{E}$. In a probabilistic graphical model, each node $i\in\mathcal{V}$ denotes a random variable $X_i$ and an edge $(i,j)\in \mathcal{E}$ is absent if and only if $X_i$ is independent of $X_j$ conditional on the other variables on the graph, $\bm{X}_{-ij}$. A clique $C$ is a subset of $\mathcal{V}$ where each pair of nodes is connected by an edge and let the set of all cliques of $\mathcal{G}$ be denoted by $\mathcal{C}$. The Hammersley--Clifford Theorem \citep{lauritzengraphical}  then relates such a graphical structure to a factorization of $f(\bm x)$ over $\mathcal{C}$ given by
$f(\bm x)=\prod_{C\in\mathcal{C}}\psi_C(\bm x_C)$, where $\psi_C(\cdot)$ is called a compatibility function, defined on a clique $C$. A graph $\mathcal{G}$ is called decomposable if it can be written as a union of edge-disjoint subgraphs, and then, the above factorization can be written in terms of marginal densities as $f(\bm x)=\prod_{C\in\mathcal{C}} f_C(\bm x_C) / \prod_{D\in\mathcal{D}}f_D(\bm x_D)$, where $\mathcal{D}$ is a multiset containing intersections between the cliques called separator sets. More details can be found in \cite{lauritzengraphical}, and Section 2.3 and Appendix A of \cite{engelke2020graphical}.

Following \cite{engelke2020graphical}, we here first consider graphs $\mathcal{G} = (\mathcal{V,E})$ that are decomposable and connected (trees in particular), with clique set $\mathcal{C}$ and separator set $\mathcal{D}$, where all separators in $\mathcal{D}$ are single nodes; thus, for $\bm{X}$ as in Section~\ref{subsub:mgpd}, $f_D(\cdot)$ denotes the density of the univariate unit-Pareto distribution, i.e., $f_D(x_D) = x_D^{-2}$ for each $D \in \mathcal{D}$. The setting of decomposable graphs is computationally attractive but overly restrictive for spatial data, and we generalize it using tree-mixtures in Section~\ref{subsub:final_model}. For each clique $C\in\mathcal{C}$, we can define a parametric MPD with intensity  $\lambda_{C}(\cdot;\bm{\theta}_C)$. 
Overall, the density over the whole graph $\mathcal{G}$ can be expressed as
\begin{equation} \label{eq:density_hr_graph}
    f_{\bm{X}}(\bm{x};\bm{\theta})=\dfrac{1}{V(\mathbf{1};\bm{\theta})}\prod_{C\in\mathcal{C}}\dfrac{\lambda_C(\bm{x}_C;\bm{\theta}_C)}{\prod_{j\in C}x_j^{-2}}\prod_{i \in \mathcal{V}} x_i^{-2},
\end{equation}
where $\bm{x}_C$ is the sub-vector of $\bm{X}$ corresponding to the clique $C$ and  $\bm{\theta} = [\bm{\theta}_C]_{C \in \mathcal{C}}$ its parameters.

\begin{figure}[t!]
\centering
\includegraphics[width=.8\linewidth]{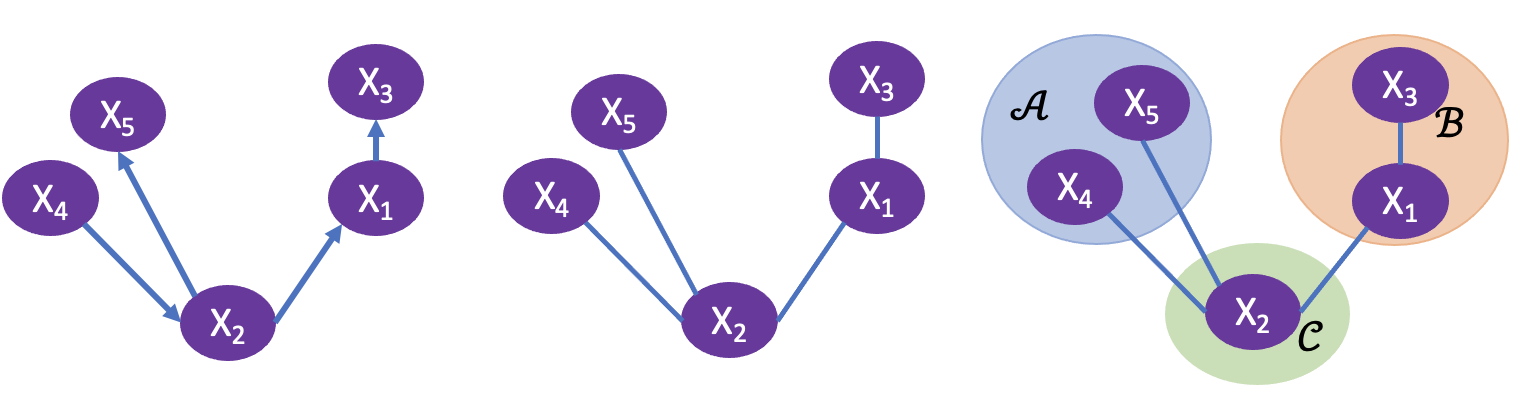} 
\caption{A directed graph (left) and the corresponding undirected graph (middle). In the right panel, for the graph in the middle, the separator set $\bm X_{\mathcal{C}}$ is highlighted in green, and $\bm X_{\mathcal{A}}$  and $\bm X_{\mathcal{B}}$ are highlighted in blue and orange, respectively.}
\label{fig::CondACondB}
 \vspace{-4mm}
\end{figure}

While the density in \eqref{eq:density_hr_graph} factorizes over a graph up to the normalizing factor $V(\mathbf{1};\bm{\theta})$, the resulting MPD random vector $\bm{X}=(X_i)_{i\in \mathcal{V}},\mathcal{V}=\{1,\ldots,N\}$, with support space $\mathcal{L}$, does not comply with the classical notion of conditional independence because $\mathcal L$ is not a product space. However, $\bm{X}$ still admits conditional independence relationships in the following modified sense. For any $k\in \mathcal{V}$, we can define the random vector $\bm{X}^{(k)}$ as $\bm{X}$ conditioned on the event $\{X_k>1\}$. Then, $\bm{X}^{(k)}$ has support on the product space $\mathcal{L}^{(k)}:=\{\mathbf{x}\in\mathcal{L}:x_k>1\}$ and has density
 $$f^{(k)}_{\bm{X}}(\bm{x})=\dfrac{f_{\bm{X}}(\bm{x})}{\int_{\mathcal{L}^{(k)}}f_{\bm{X}}(\bm{x}) d\bm{x}},~ \bm{x} \in \mathcal{L}^{(k)},$$
where $f_{\bm{X}}(\cdot)$ is as in Section~\ref{subsub:mgpd}. Let $\mathcal{A},\mathcal{B},\mathcal{C} \subset \mathcal{V}$ be non-empty disjoint subsets whose union is $\mathcal{V}$, as shown in the right panel of Figure~\ref{fig::CondACondB}, and also let the corresponding components of $\bm{X}$ be $\bm{X}_{\mathcal{A}}$, $\bm{X}_{\mathcal{B}}$, and $\bm{X}_{\mathcal{C}}$, respectively. Similarly, let also $\bm{X}^{(k)}_{\mathcal{A}}$, $\bm{X}^{(k)}_{\mathcal{B}}$, and $\bm{X}^{(k)}_{\mathcal{C}}$ be the respective components of $\bm{X}^{(k)}$. Then, \cite{engelke2020graphical} show that graphical structure of $\bm{X}$ is equivalent to the (classical) conditional conditional independence $\bm X^{(k)}_{\mathcal{A}} \indep \bm X^{(k)}_{\mathcal{B}}\mid\bm X^{(k)}_{\mathcal{C}}$ for all $k \in \mathcal{V}$, and in this case, they write $\bm X_{\mathcal{A}} \bot_e \bm X_{\mathcal{B}} \mid \bm X_{\mathcal{C}}$ to emphasize that this new notion adapted for multivariate extremes.



A tree $\mathcal{T}=(\mathcal{V,E})$ with nodes $\mathcal{V}$ and edges $\mathcal{E}$ is a connected undirected graph without cycles, and cliques of maximum dimension two, and thus, the relation between the cardinalities of $\mathcal{V}$ and $\mathcal{E}$ is $|\mathcal{E}|=|\mathcal{V}|-1$. In Figure~\ref{fig::CondACondB}, all the graphs are trees as there is a unique path between any two nodes; the first graph (left) is a directed tree and the second (middle) graph is an undirected tree. If $\bm X$ is an extremal graphical model satisfying the global Markov property with respect to a tree $\mathcal{T}$, we obtain a simple stochastic representation for $\bm X^{(k)}$. Its extremal variogram rooted at node $k\in\mathcal{V}$ is defined as the matrix $\bm{\Gamma}^{(k)}$ with its ${(i,j)\mbox{-th}}$ entry is given by $\Gamma^{(k)}_{ij}=\textrm{Var}\{\log X_i^{(k)}-\log X_j^{(k)}\},i,j\in \mathcal{V}$, whenever the right-hand side is finite. Here $\Gamma^{(k)}_{ij}$ turns out to be an additive tree metric, and thus it is a natural quantity to infer conditional independence in extremal tree models. \cite{engelke2020structure} define the extremal variogram for arbitrary MPDs, and in the special case of the H\"usler--Reiss distribution with variogram matrix $\bm \Gamma$, they show that $\Gamma_{ij}=\Gamma_{ij}^{(k)}$ for all $k=1,\ldots,N$, such that for tree-based H\"usler--Reiss MPD models, 
\begin{eqnarray}\label{eq:Gamma_ij}
\Gamma_{ij}=\sum_{(i^*,j^*)\in \mathcal{P}(i,j;\mathcal{T})}\Gamma_{i^*j^*},\qquad k=1,\ldots,d,
\end{eqnarray}
where $\mathcal{P}(i,j;\mathcal{T})$ denotes the set of edges on the unique path from node $i$ to node $j$ on tree $\mathcal{T}$. 

\subsubsection{Final model based on tree-mixtures}
\label{subsub:final_model}

Our focus is on extending the extremal graphical modeling methodology of \cite{engelke2020graphical} to (potentially nonstationary) spatial extremes. Graph decomposability, a crucial assumption in Section~\ref{subsub:single_tree}, is overly restrictive for spatial data, and tree structures in non-trivial settings do not allow spatial stationarity or realistic nonstationarity on regular lattice data in $\mathbb{R}^2$. The most natural spatially-structured Markov random field on a regular lattice does not satisfy decomposability, making it difficult to extend the methodology to this case. \cite{yu2016modeling} and \cite{vettori2020bayesian} propose using finite mixtures of tree-based models to retain computational tractability, and we build upon their approaches in the context of spatial extremes using tree-based MPDs. 

Here, we assume that the dependence structure can be explained through a mixture of $M$ tree-based MPD models. For a general undirected graph $\mathcal G=(\mathcal{V},\mathcal{E})$, let the set of all spanning trees (thus, subgraphs of $\mathcal G$) be denoted by $\widetilde{\mathcal{T}} = \{\mathcal{T}_1, \ldots, \mathcal{T}_M \}$. While the nodes for all such trees are the same as they represent the same data locations, the set of edges are different, and we thus denote the trees by $\mathcal{T}_m = (\mathcal{V}, \mathcal{E}_m )$, $m=1,\ldots,M$. Since trees have the property that all cliques are of maximum dimension two and separators contain only one node, the MPD density based on $\mathcal{T}_m$ is 
\begin{align}\label{eq:join_tree}
f_{\bm{X}}(\bm x\mid \mathcal{T}_m; \bm{\theta}) = \dfrac{1}{V(\mathbf{1};\bm{\theta})}\prod_{ \{ i,j \} \in \mathcal{E}_m}\dfrac{\lambda_{ij}(x_i, x_j;\bm\theta_{ij})}{x_i^{-2} x_j^{-2}}\prod_{i \in \mathcal{V}} x_i^{-2},
\end{align}
where $\lambda_{ij} \equiv \lambda_{(i,j)}$ are the bivariate marginals of the exponent measure density $\lambda(\cdot;\bm \theta)$ and $\bm\theta_{ij}$ is the dependence parameter vector corresponding to the edge $(i,j)\in\mathcal E$. For the specific H\"usler--Reiss model, $\lambda_{ij}$ can be deduced from \eqref{eq:lambda_hr}, while the edge parameter is simply equal to $\bm\theta_{ij}=\Gamma_{ij}$, the ${(i,j)\mbox{-th}}$ element of the matrix $\bm\Gamma$. We construct our model based on an ensemble-of-tree mixture as 
\begin{align}\label{eq:ETPHR}
\widetilde{f}_{\bm{X}}\left(\bm x; \bm{\theta}\right) = \sum_{m=1}^{M} p(\mathcal{T}_m)~f_{\bm{X}}\left(\bm x\mid \mathcal{T}_m; \bm{\theta}\right),
\end{align}
where $p(\mathcal{T}_m)$'s are tree probabilities assigned by the decomposable prior over spanning trees proposed by \cite{meilua2006tractable}, which takes the specific form 
\begin{align}\label{eq:prior_t}
p(\mathcal{T}_m)= \dfrac{1}{C_{\widetilde{\mathcal{T}}}}\prod_{(i,j)\in\mathcal{E}_m}\beta_{ij}, ~~m=1,\ldots,M,
\end{align} 
where $C_{\widetilde{\mathcal{T}}} = \sum_{m=1}^M \prod_{(i,j)\in\mathcal{E}_m}\beta_{ij}$, and $\beta_{ij}$ are edge weights to be computed. We denote the edge weight matrix by $\bm{\beta}$ with its ${(i,j)\mbox{-th}}$ element being $\beta_{ij}$. Here, $\bm{\beta}$ is symmetric and its diagonal entries are zero. Following Kirchhoff's matrix theorem \citep{beineke2004topics}, we have $C_{\mathcal{T}} = \textrm{det}\{\bm{Q}(\bm\beta)\}$,
where $\bm{Q}(\bm\beta)$ denotes the first $(N-1)$ rows and columns of the $(N \times N)$-dimensional Laplacian matrix $\bm{L}(\bm \beta)= \textrm{diag}(\bm \beta \bm 1_N)-\bm \beta$ corresponding to a graph with edge weight matrix $\bm{\beta}$. Here $\bm 1_N$ is a $N$-length column vector of all ones. By substituting \eqref{eq:prior_t} with  $C_{\mathcal{T}} = \textrm{det}\{\bm{Q}(\bm\beta)\}$ into \eqref{eq:ETPHR}, we have 
\begin{align}\label{eq:final_ETH}
\widetilde{f}_{\bm{X}}\left(\bm x; \bm{\theta}\right) = \dfrac{1}{\textrm{det}\{\bm{Q}(\bm\beta)\} V(\mathbf{1};\bm{\theta})} \prod_{i \in \mathcal{V}} x_i^{-2} \sum_{m=1}^{M} \prod_{ \{ i,j \} \in \mathcal{E}_m} \beta_{ij} \dfrac{\lambda_{ij}(x_i, x_j;\bm\theta_{ij})}{x_i^{-2} x_j^{-2}}.
\end{align}
Further, by denoting $\lambda^*_{ij}(x_i, x_j;\bm\theta_{ij}) = \lambda_{ij}(x_i, x_j;\bm\theta_{ij}) / (x_i^{-2} x_j^{-2}) $, we obtain
\begin{align}\label{eq:final_ETH}
\widetilde{f}_{\bm{X}}\left(\bm x; \bm{\theta}\right) = \dfrac{1}{V(\mathbf{1};\bm{\theta})} \prod_{i \in \mathcal{V}} x_i^{-2} \times \dfrac{\textrm{det}\{\bm{Q}(\bm\beta \odot \bm{\lambda}^*)\}}{\textrm{det}\{\bm{Q}(\bm\beta)\}},
\end{align}
where $\bm{\lambda}^*$ is a $(N\times N)$-dimensional matrix with its ${(i,j)\mbox{-th}}$ element being $\lambda^*_{ij}(x_i, x_j;\bm\theta_{ij})$, $\odot$ denotes element-wise multiplication of two matrices, and $\bm{Q}(\cdot)$ denotes the same matrix operation as above. 




Once the weights $\bm\beta$ are obtained (see Section~\ref{subsec:computebeta}), we can then compute the prior probabilities of each spanning tree in \eqref{eq:prior_t} and combine these tree-based MPD models as in \eqref{eq:ETPHR} to construct the full mixture model. However, there are a number of drawbacks in building this mixture using all possible spanning trees. First, depending on the initial graph $\mathcal G$, the number of spanning trees can be huge, especially if the number of nodes $|\mathcal{V}|$ (i.e., number of sites in spatial applications) is large or if $\mathcal G$ is densely connected, making it computationally impractical. Second, not all these spanning trees might be meaningful in a spatial context, like the one we are interested in. While the second drawback can be somewhat mitigated by choosing the initial graph appropriately (e.g., as a simple lattice), the first drawback remains problematic in high-dimensional spatial settings. Therefore, instead of considering all possible spanning trees $\widetilde{\mathcal{T}}$ as in \eqref{eq:ETPHR}, we propose setting some prior tree probabilities to zero and constructing the tree-mixture from a smaller sub-selection of suitably sampled random trees. We set the number of such random trees to be quite large, but typically much smaller than $M=|\widetilde{\mathcal{T}}|$, which makes it computationally feasible, and then compute prior weights as in \eqref{eq:prior_t} but with a different normalizing factor to account for the fact that a large number of (unsampled) trees have prior probability zero.

There are various ways to generate random trees. Here, we proceed as follows to sample a variety of random trees that can collectively represent the spatial dependence structure of the data appropriately. Assuming that the initial graph $\mathcal G=(\mathcal V,\mathcal E)$ is a regular lattice (which is the case in our application), we first draw some random positive values $\omega_{ij}$ assigned to all edges $(i,j)\in\mathcal E$; we need to make sure that $\omega_{ji}=\omega_{ij}$ and $\omega_{ii}=0$, but these values do not need to be the same for the different edges (in particular different edge types). For example, in our case, we draw these random values differently depending on whether the edge between two sites is horizontally (W--E), vertically (N--S), or diagonally (NW--SE and SW--NE) oriented on the lattice $\mathcal G$. This allows favoring trees along certain directions, which allows capturing some form of anisotropy and nonstationarity. Then, we compute the minimum spanning tree, which minimizes a weighted sum of the sampled random edge values $\omega_{ij}$, and the computed edge weights $\beta_{ij}$. The minimum spanning tree is unique once the edge values have been sampled. However, by repeated sampling, our procedure can therefore generate a wide range of different random trees whose edges follow certain desired patterns.

\subsubsection{Model properties}

Given a tree $\mathcal{T}_m$, the marginal distribution for each component $X_i$ of the MPD vector $\bm{X}$ in \eqref{eq:join_tree}, given that $X_i>1$, is unit-Pareto with density $x_i^{-2}, x_i \geq 1$, which does not rely on the dependence parameter vector $\bm{\theta}$. Therefore, the same holds for the tree-mixture in \eqref{eq:ETPHR}. 
Hence, our proposed tree-mixture model is effectively a copula model (on a scale different from uniform), which concerns the modeling of the extremal dependence structure. 

For multivariate extremes, several summary statistics have been developed to measure the strength of dependence between the extremes of two variables $X_i$ and $X_j$ \citep{coles1999dependence}. The most popular one is the $\chi$-measure (recall its definition in Section~\ref{sec:DataDescr}),
expressed as $\chi_{ij}=\lim_{u\rightarrow 1}\textrm{P}\{F_i(X_i)>u\mid F_j(X_j)>u\}$ whenever the limits exists, where $F_i$ and $F_j$ denote the marginal distribution functions of $X_i$ and $X_j$, respectively. 

For a MPD random vector $\bm X$ defined in \eqref{cdf_exceedance}, the $\chi$-measure between the components $X_{i}$ and $X_{j}$ can be expressed as $\chi_{ij}=2-V_{ij}(1,1)$, where $V_{ij}$ denotes the underlying exponent function restricted to the subvector $(X_{i},X_{j})'$. In particular, for the H\"usler--Reiss distribution with variogram matrix $\bm\Gamma$, which is closed under marginalization, we get $\chi_{ij}=2\Bar{\Phi}(\sqrt{\Gamma_{ij}/2})$, where $\Bar{\Phi}(\cdot)$ denotes the standard normal survival function and $\Gamma_{ij} = \textrm{Var}\{\log(X_i) - \log(X_j)\}$ is the ${(i,j)\mbox{-th}}$ element of $\bm\Gamma$. In case of a tree-based H\"usler--Reiss distribution defined over a single tree $\mathcal{T}_m = (\mathcal{V},\mathcal{E}_m)$, we thus get from \eqref{eq:Gamma_ij} that for any $i,j\in\mathcal{V}$,
\begin{align}
\label{eq:Chi_tree}
\chi_{ij}&=2\Bar{\Phi}\left(\sqrt{\Gamma_{ij}}\right)=2\Bar{\Phi}\left(\sqrt{\sum_{(i^*,j^*)\in \mathcal{P}(i,j;\mathcal{T}_m)}\Gamma_{i^*j^*}}\right),
\end{align}
where $\mathcal{P}(i,j;\mathcal{T}_m)$ denotes the set of edges on the unique path from node $i$ to node $j$ on tree $\mathcal{T}_m$. Thus, for our final tree-mixture spatial model $\widetilde{f}_{\bm{X}}(\cdot)$ in \eqref{eq:final_ETH}, we deduce from \eqref{eq:Chi_tree} and Equation (61) in Lemma~1 of \citet{yu2016modeling} that 
the $\chi$-measure equals
\begin{align}
\label{eq:Chi_final}
\chi_{ij} &=\lim_{u\rightarrow 1}\sum_{m=1}^M p(\mathcal{T}_m)\;\textrm{P}\{F_{i}(X_{i})>u\mid F_{j}(X_{j})>u,\mathcal{T}_m\}\nonumber\\
&=2\sum_{m=1}^M p(\mathcal{T}_m)\;\Bar{\Phi}\left(\sqrt{\sum_{(i^*,j^*)\in \mathcal{P}(i,j;\mathcal{T}_m)}\Gamma_{i^*j^*}}\right),
\end{align}
where $p(\mathcal{T}_m)$ is defined as in \eqref{eq:prior_t}. Note that the prior probabilities $p(\mathcal{T}_m)$ are properly modified when fewer trees are used, as explained in Section~\ref{subsub:final_model}. Unlike \citet{yu2016modeling}, who only derive lower bounds for the $\chi$-measure, $\chi_{ij}$ can in our case be obtained in closed form. Note that in the case of spatial data on a 2D lattice $\mathcal G=(\mathcal V,\mathcal E)$, the dependence structure from a single-tree H\"usler--Reiss MPD model in \eqref{eq:Chi_tree} is always (artificially) non-stationary, while that from the tree-mixture model in \eqref{eq:Chi_final} is \emph{in general} non-stationary, but it can also lead to stationarity depending on the model parameters and the prior tree probabilities. It is therefore a much richer class of models.

\section{Inference}
\label{sec:computation}


Marginal generalized Pareto distribution parameters are estimated by maximizing the likelihood function \eqref{neighboring_like}. Estimates are then used to transform the data to the unit-Pareto scale, as illustrated in Figure~\ref{fig:flowchart_stage1}. In this section, we consider the tree-based MPD mixture \eqref{eq:join_tree} with trees supported over an initial graph $\mathcal G=(\mathcal V,\mathcal E)$ and MPD mixture components of H\"usler--Reiss type parametrized by the variogram matrix $\bm\Gamma$; we first explain how to estimate $\bm \Gamma$ and then show how to compute the edge weights $\bm\beta$ in \eqref{eq:prior_t} used in the final model density \eqref{eq:final_ETH}. The Supplementary Material contains a simulation study that illustrates the good performance of our estimation approach.

\subsection{Estimation of $\boldmath{\Gamma}$}
\label{sub:theta_est}

From \eqref{eq:join_tree}, it is clear that for a MPD constructed from a tree $\mathcal T_m=(\mathcal V,\mathcal E_m)$ spanning the graph $\mathcal G=(\mathcal V,\mathcal E)$, the only relevant parameters in $\bm \Gamma$ that need to be estimated are those that correspond to all edges $\mathcal E_m$ in the tree $\mathcal T_m$ (which are the only cliques). Given that our final model is a mixture of these spanning tree-based MPDs, we need to estimate parameters $\Gamma_{ij}$ over all edges of the initial graph $\mathcal G$. While it would be possible to impose some ``spatial structure'' to the parameters, e.g., by forcing all similar edge types (e.g., horizontal, etc.) to share the same parameter value across the graph, which would mimic stationarity, we have found that this solution is suboptimal in our case, and we thus estimate parameters $\Gamma_{ij}$ edge-by-edge for all $(i,j)\in\mathcal E$. This approach leads to a less parsimonious but more flexible model that can capture complex forms of spatial nonstationarity. 

Suppose $\bm{Y}_{i,j,t} = (Y_{i,t}, Y_{j,t})'$ denotes the bivariate vector of standardized Pareto observations corresponding to pixels $i$ and $j$ at time $t$, where the $Y_{i,t}$'s are obtained according to the marginal transformation procedure outlined in Section~\ref{sub:Stage1}. The corresponding limiting MPD that arise for high threshold exceedances according to \eqref{cdf_exceedance} is here assumed to follow a bivariate H\"usler--Reiss distribution; recall \eqref{eq:lambda_hr}. We thus fit the bivariate H\"usler--Reiss distribution to bivariate exceedances from $\bm{Y}_{i,j,t}$ such that $\|\bm{Y}_{i,j,t}\|_{\infty}>u$ for some high threshold $u>1$ using a pairwise censored likelihood. From \eqref{cdf_exceedance}, the contribution of the ${t\mbox{-th}}$ time point to the censored likelihood is
\begin{align*} 
L_t(\Gamma_{ij})=\begin{cases}
\lambda(Y_{i,t}/u,Y_{j,t}/u;\Gamma_{ij})/V(1,1;\Gamma_{ij}),&Y_{i,t} > u, Y_{j,t} > u,\\
-V_1(Y_{i,t}/u,1;\Gamma_{ij})\}/V(1,1;\Gamma_{ij}),&Y_{i,t} > u, Y_{j,t} \leq u,\\
-V_2(1,Y_{j,t}/u;\Gamma_{ij})/V(1,1;\Gamma_{ij}),&Y_{i,t} \leq u, Y_{j,t} > u,
\end{cases}
\end{align*}
where $V_k(x_i,x_j;\Gamma_{ij})=\partial V(x_i,x_j;\Gamma_{ij})/\partial x_k$, $k=i,j$, and $\lambda(x_i,x_j;\Gamma_{ij})=-\partial^2 V(x_i,x_j;\Gamma_{ij})/(\partial x_i\partial x_j)$. Then, assuming independence over $t$, the full censored log-likelihood is 
$\ell(\Gamma_{ij})=\sum_{t=1}^T\log\{L_t(\Gamma_{ij})\},$ 
and the corresponding parameter estimate is obtained as $\hat{\Gamma}_{ij} = \argmax_{\Gamma_{ij}>0} \ell(\Gamma_{ij})$. This optimization needs to be repeated for all edges $(i,j)\in\mathcal{E}$. Explicit forms of the censored likelihood for parametric models other than H\"usler--Reiss can be found, for example, in \citet{coles2001introduction}.  

\subsection{Computation of $\boldmath{\beta}$}
\label{subsec:computebeta}

Once we the parameters for each edge $(i,j)\in\mathcal{E}$ of the initial graph $\mathcal{G}=(\mathcal{V},\mathcal{E})$ are estimated, we build the mixture of tree-based MPDs in an optimal way by finding the edge weights $\beta_{ij}$ in \eqref{eq:ETPHR} that maximize the overall mixture likelihood; in other words, we adjust the prior tree probabilities $p(\mathcal{T}_m)$ in order to provide the best fit to the data. This is done by leveraging the expression  \eqref{eq:final_ETH} and thus computing the edge weight matrix $\bm\beta$ as 
\begin{align}
\label{eq:maxBeta}
\hat{\bm \beta}=\argmax_{\bm \beta} \left\lbrace \sum_{t=1}^{T}\log \textrm{det}[Q(\bm \beta\odot \hat{\bm L}_t)]-T\log\textrm{det }[Q(\bm \beta)] \right \rbrace,
\end{align}
such that $\beta_{ij}\geq 0$ for all $(i,j)\in\mathcal{E}$ (with $\beta_{ij}=0$ for all $(i,j)\notin\mathcal{E}$) and $\|U(\bm \beta)\|_2=1$, where $\|U(\cdot)\|_2$ is the Euclidean norm for the upper triangular part of the input matrix, while $\hat{\bm L}_t$ is the corresponding matrix of estimated pairwise censored likelihood contributions for the ${t\mbox{-th}}$ replicate (i.e., the values $L_t(\hat\Gamma_{ij})$ for each edge $(i,j)\in\mathcal{E}$, arranged in matrix form). For optimizing the likelihood in \eqref{eq:maxBeta}, we use a doubly-stochastic gradient-ascent (DSGA) algorithm, similar to \citet{yu2016modeling}.

The optimization problem in \eqref{eq:maxBeta} is challenging due to the constraints on $\bm \beta$ and the complexity of the likelihood function. To ease the optimization, we relax the constraints in \eqref{eq:maxBeta} and instead solve the regularized maximization problem
\begin{align}
\label{eq:maxBeta2}
    \hat{\bm \eta}=\argmax_{\bm \eta=\log\{\bm\beta\}} \left\lbrace \sum_{t=1}^{T} \log \det [Q(\bm \beta\odot \hat{\bm L}_t) ]-T \log\textrm{det }[Q(\bm \beta)]-a_1(\|U(\bm \beta)\|_2^2-1)^2 \right \rbrace,
\end{align}
for some fixed positive constant $a_1$ and where $\eta_{ij}=\log\{\beta_{ij}\}$ for all $(i,j)\in\mathcal{E}$. The DSGA algorithm is a type of stochastic gradient ascent algorithm that updates the parameter vector $\bm\eta$  in small steps based on the gradients of the log-likelihood function with respect to $\bm \eta$. 
 

The stochastic gradient descent algorithm involves initializing the parameter vector $\bm{\eta}$ and choosing a mini-batch of data denoted by $\mathcal{B}$; the algorithm then proceeds by iteratively updating the parameter vector along the gradient of the log-likelihood (see Algorithm \ref{algorithm_sgd} for details). The DSGA algorithm has two additional features that improve its convergence. First, it adds a row and column-normalization step after each update which ensures the resulting matrix to have unit norm. Second, it uses a step size that decreases over time to ensure convergence.


\begin{algorithm}[t!]
\caption{Stochastic Gradient Descent Algorithm}
\begin{algorithmic}[1]
\State Initialize the parameter vector $\bm{\eta}$ as $\bm{\eta}^{\{0\}}$. Set iteration number to $k=1$.
\State Choose a mini-batch of the data, denoted by $\mathcal{B}$.
\State Compute the gradient of the log-likelihood function \eqref{eq:maxBeta2} with respect to $\bm{\eta}$ using only the data in the mini-batch, denoted by $\nabla_{\bm\eta}g$.
\State Update the parameter vector using the gradient as $\bm{\eta}^{\{k\}}=\bm{\eta}^{\{k-1\}}+\rho^{\{k-1\}}\tilde{\nabla}_{\bm\eta}g|_{\bm\eta=\bm\eta^{\{k-1\}}}$, where $\rho^{\{k-1\}}$ is the current step size.
\State Project the updated parameter vector onto the feasible set, which consists of matrices with non-negative symmetric entries and unit upper-triangular Euclidean norm.
\State Update iteration number to $k\mapsto k+1$
\State Repeat steps 2--6 until convergence.
\end{algorithmic}
\label{algorithm_sgd}
\end{algorithm}

\subsection{Bias-corrected extremal dependence estimate}
\label{sub:chi_correction}

Once the dependence parameters in $\bm\Gamma$ are estimated and the edge weights in $\bm\beta$ are computed, the estimated extremal dependence structure for our mixture model of tree-based MPDs can be easily obtained from \eqref{eq:Chi_tree}. However, in spatial applications, trees are only an approximate representation of reality and when the data are truly stationary and isotropic we have found that the tree-mixture model tends to underestimate the extremal dependence structure, especially at large distances; this is due to the fact that the dependence between two sites in the tree-mixture model must follow the paths between these two nodes (on the trees considered in the mixture), which are usually longer than their actual Euclidean spatial distance. This distortion of distances inevitably leads to an underestimation of the dependence strength. In order to mitigate this issue, we propose a simple post-processing step, where the estimated variogram matrix $\hat{\bm\Gamma}$ is scaled by a factor $a\in(0,1]$. The resulting $\chi$-measure can be computed as in \eqref{eq:Chi_tree}, but with the $\Gamma_{ij}$ values replaced with $a\hat\Gamma_{ij}$, i.e., 
\begin{align}
\label{eq:Chi_updated}
\hat{\chi}_{ij}(a) &=2\sum_{m=1}^M \hat{p}(\mathcal{T}_m)\; \Bar{\Phi}\left(\sqrt{\sum_{(i^*,j^*)\in \mathcal{P}(i,j;\mathcal{T}_m)} a \hat{\Gamma}_{i^*j^*}}\right). 
\end{align}
To estimate the optimal value of $a$, we then minimize the $L_2$-distance between the empirical $\chi_{ij}$, say $\hat{\chi}_{ij}^{\rm emp}$, and $\hat\chi_{ij}(a)$ in \eqref{eq:Chi_updated}. For $\hat{\bm\chi}^{\rm emp}=\{\hat\chi_{ij}^{\rm emp}\}_{(i,j)\in\mathcal{E}}$ and $\hat{\bm\chi}(a)=\{\hat\chi_{ij}(a)\}_{(i,j)\in\mathcal{E}}$, the optimal $a$ is thus obtained as 
$a_{\textrm{opt}} = \argmin_{a\in(0,1]}\|\hat{\bm\chi}^{\rm emp}-\hat{\bm\chi}(a)\|^2_2.$ 
This simple bias-correction procedure works quite well as illustrated in a simulation study reported in the supplementary material of this paper. 


\section{Wildfire data application} 
\label{sec:Application}


Motivated by the exploratory analysis in Section~\ref{sec:DataDescr}, we analyze the extreme FFDI data using a local moving window approach to account for possible time trends. Specifically, we split the study period of 23 summer seasons (Nov., Dec., Jan., Feb.) into 14 decadal windows and perform window-specific analyses. Here, the first and the last windows are 1999--2009 and 2012--2022, respectively.

\begin{figure}[t]
\centering
\includegraphics[width = 0.032\linewidth]{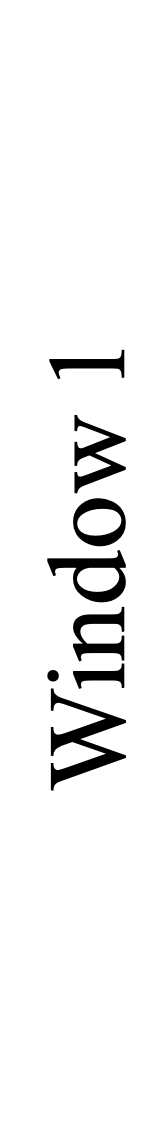}
\includegraphics[width = 0.315\linewidth]{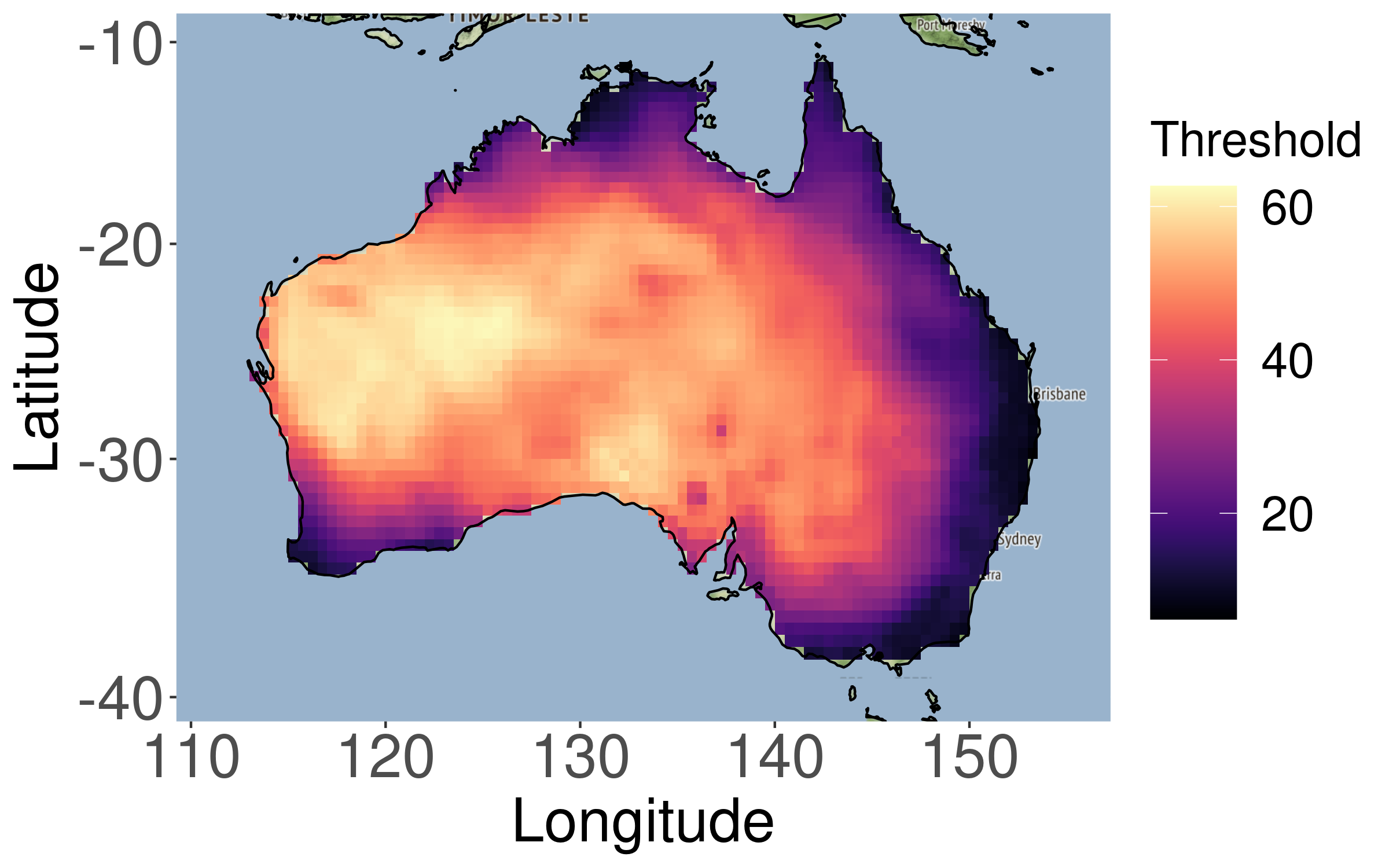}
\includegraphics[width = 0.315\linewidth]{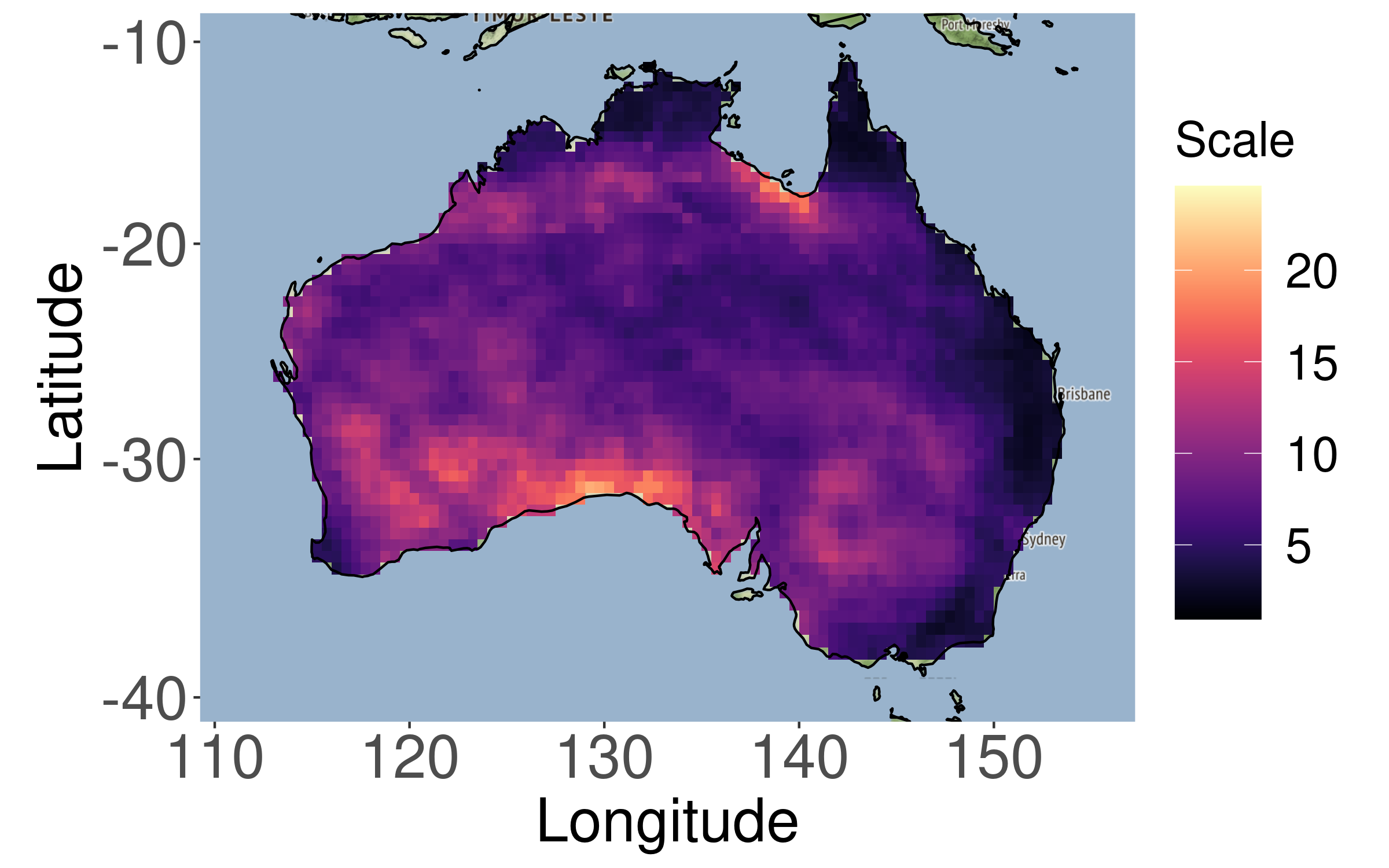}
\includegraphics[width = 0.315\linewidth]{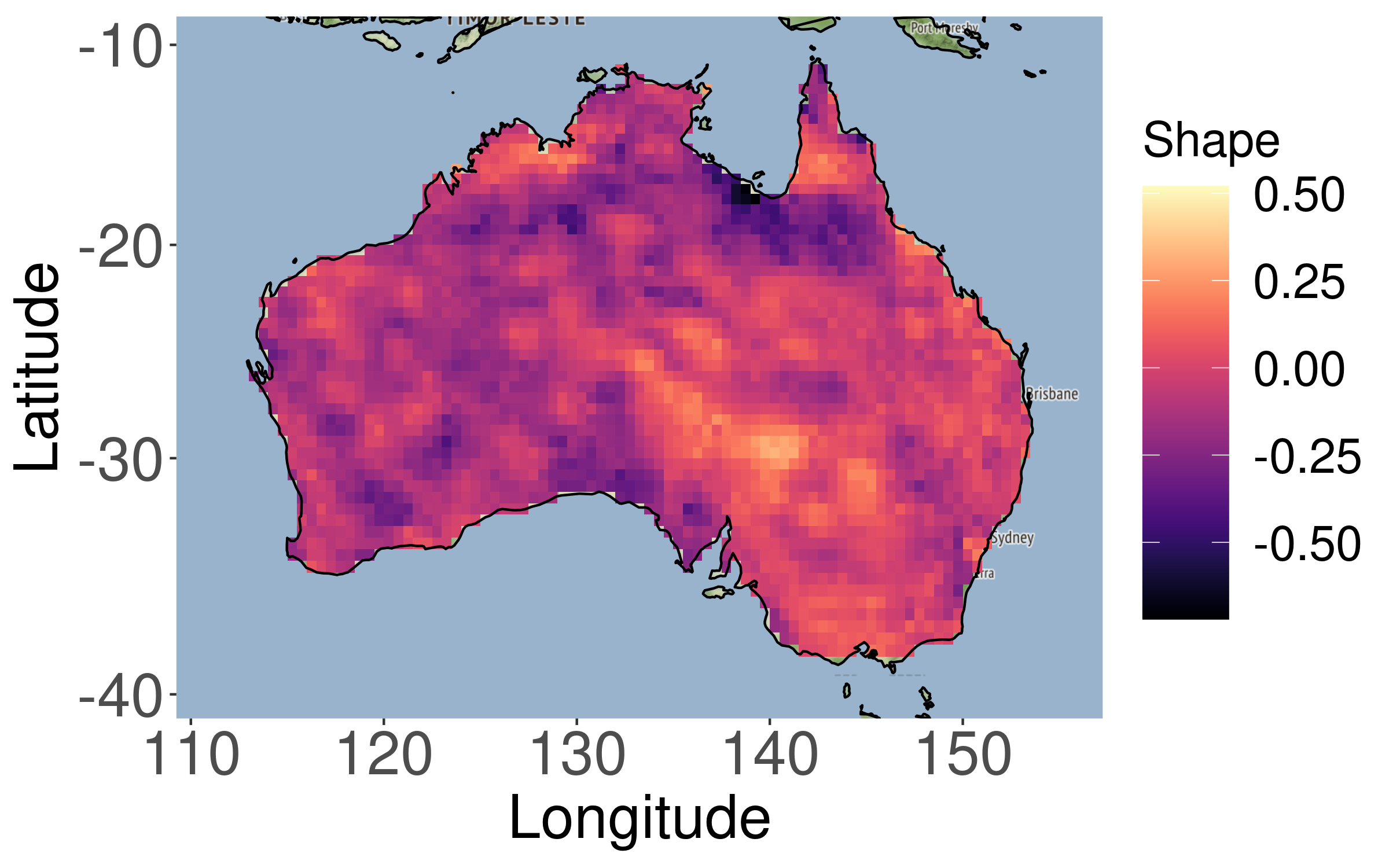}

\includegraphics[width = 0.032\linewidth]{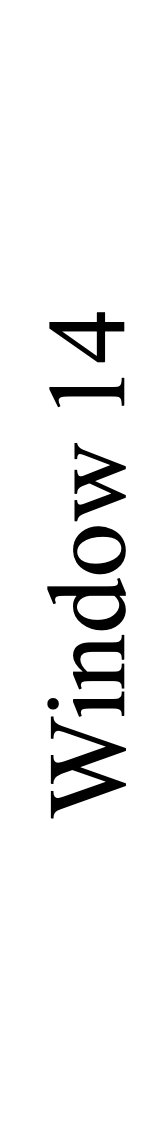}
\includegraphics[width = 0.315\linewidth]{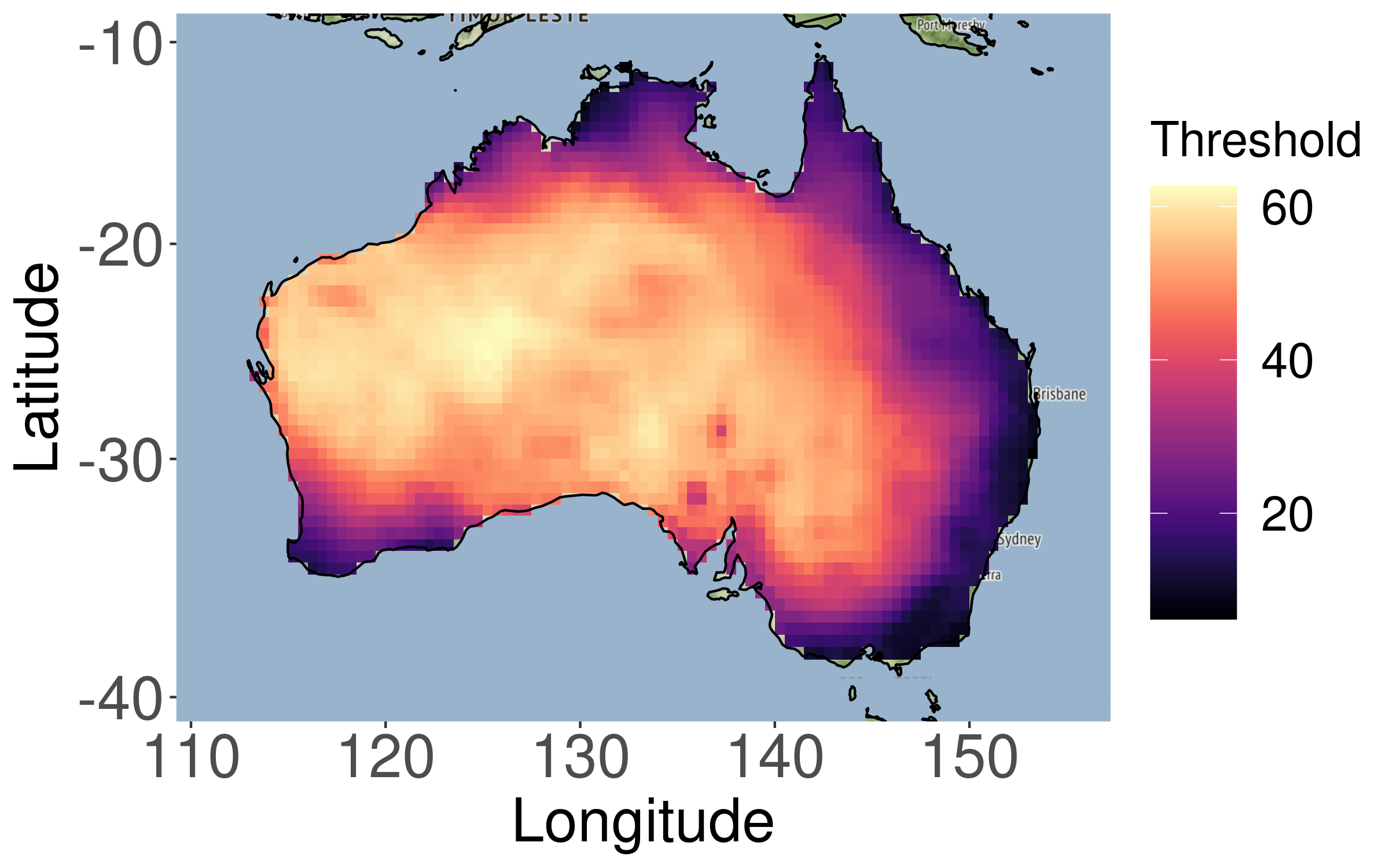}
\includegraphics[width = 0.315\linewidth]{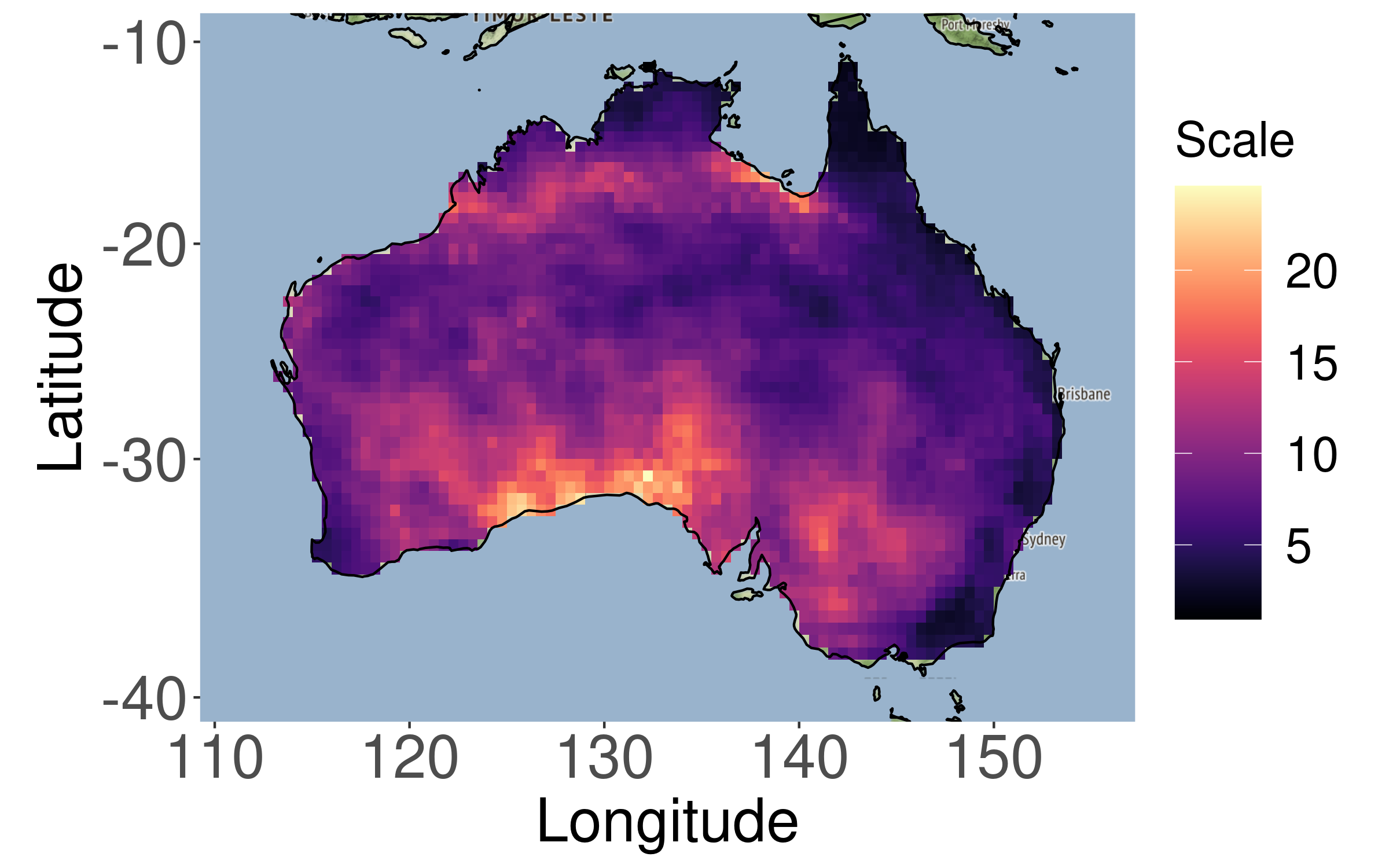}
\includegraphics[width = 0.315\linewidth]{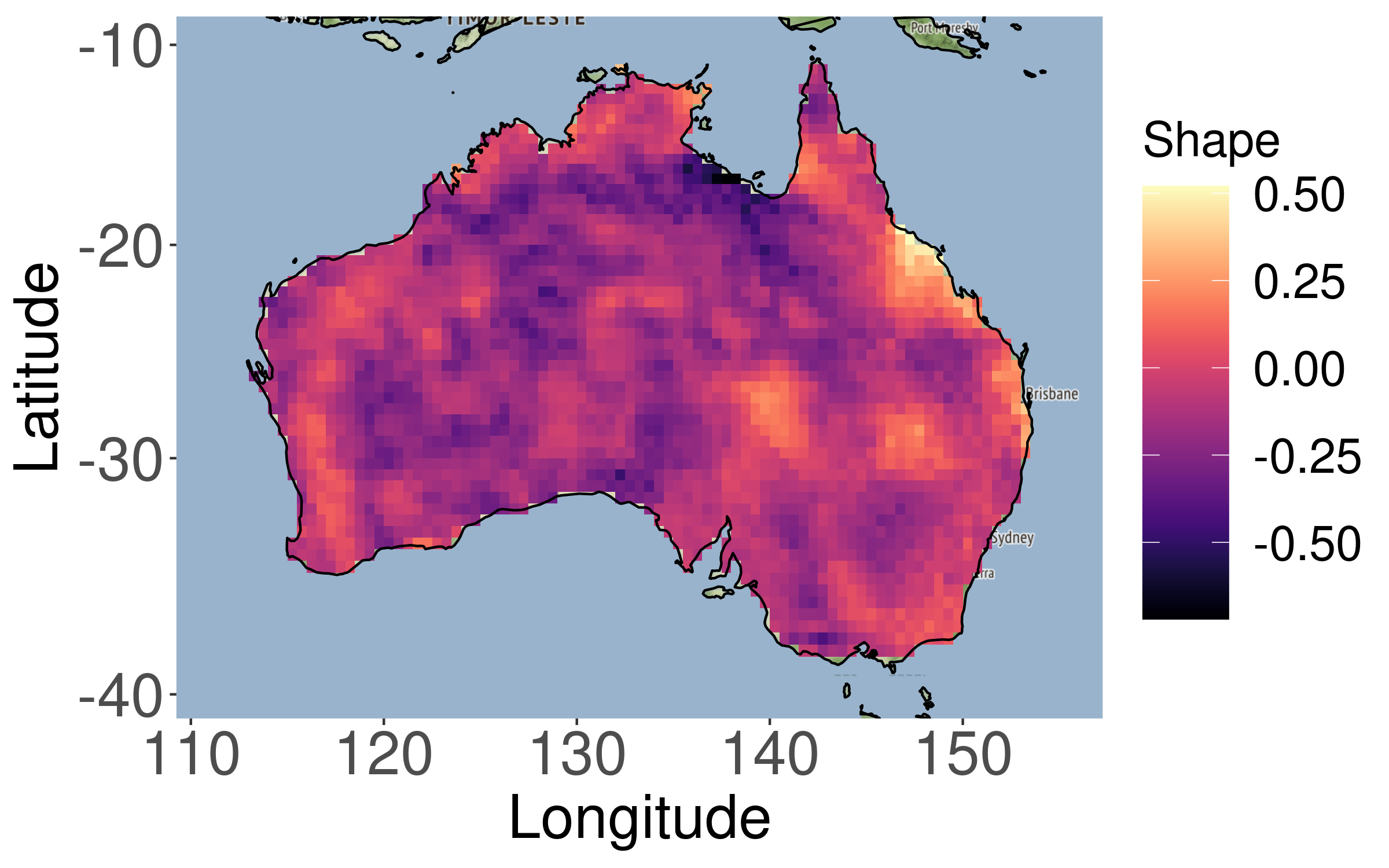}
   
\caption{Spatial maps of the estimated thresholds (left), the generalized Pareto scale (middle), and shape (right) parameters for the first (top) and the last (bottom) temporal windows.}
\label{fig:Parameters}
\vspace{-2mm}
\end{figure}

The first step of the analysis is to fit marginal distributions. We follow the procedure outlined in Section~\ref{sub:Stage1}, whereby the GP distribution is fitted locally to high threshold exceedances (here, taken to be the empirical $95\%$-quantile at each site). For estimation, we pool data from neighboring sites, which helps to control spatial smoothness in estimated parameters and to keep marginal inference numerically stable. Figure~\ref{fig:Parameters} displays spatial maps of the estimated thresholds, GP scales, and GP shape parameters, for the first and the last windows. 
The estimated thresholds are spatially smooth apart from a few ``hotspots'' near the southern coast. They show higher values near the middle and western regions of Australia and are lower near the eastern and northeastern coastal regions and a small southwestern region near the city of Perth. By contrast, the estimated GP parameters (and especially the shape) show more local variability, which is due to the relatively small sample size available to estimate these two tail parameters locally. The local data pooling approach is therefore crucial. From the maps, we can see that the GP scale parameter takes higher values near the southern coast and a small portion along the northern coast, while the GP shape parameter generally lies within the interval $[-0.25,0.25]$ and is usually close to zero or slightly negative. While it is difficult to see significant differences between the estimates for the first and the last temporal windows, it can still be noticed that the GP scale for Window 14 is higher than for Window 1 across quite a large portion of the study domain, which is indicative of a temporal shift in FFDI data across the country. To investigate this further, we computed the marginal $99\%$-quantile based on our fitted model using \eqref{eq:retlev} for both temporal windows; we found that estimated quantiles for Window 14 are higher than those for Window 1 at $1\,949$ out of $2\,749$ grid cells (i.e., about $71\%$ of the domain, including major cities), thus confirming that wildfire risk has worsened over time. 

\begin{figure}[t!]
\centering 
\includegraphics[width=.76\paperwidth]{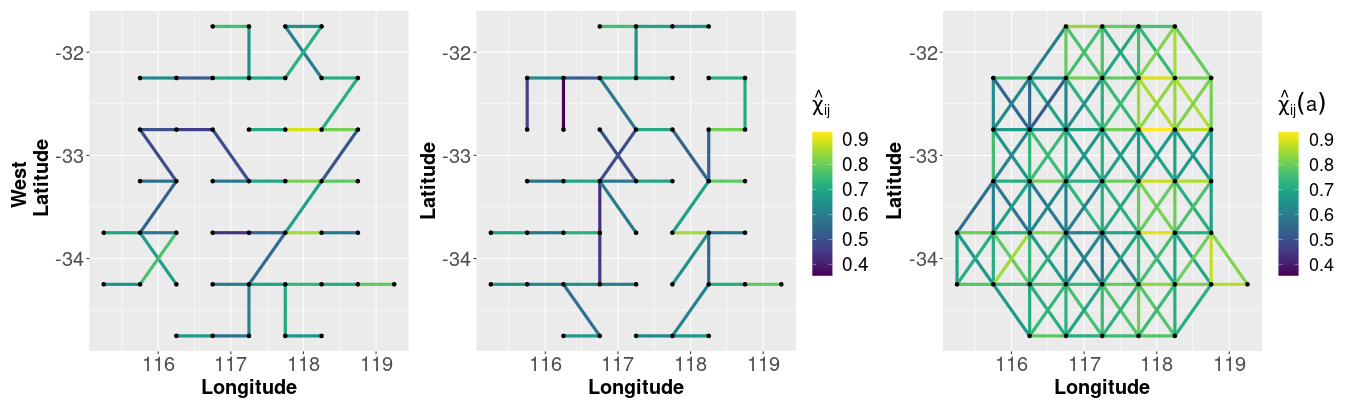}
\includegraphics[width=.76\paperwidth]{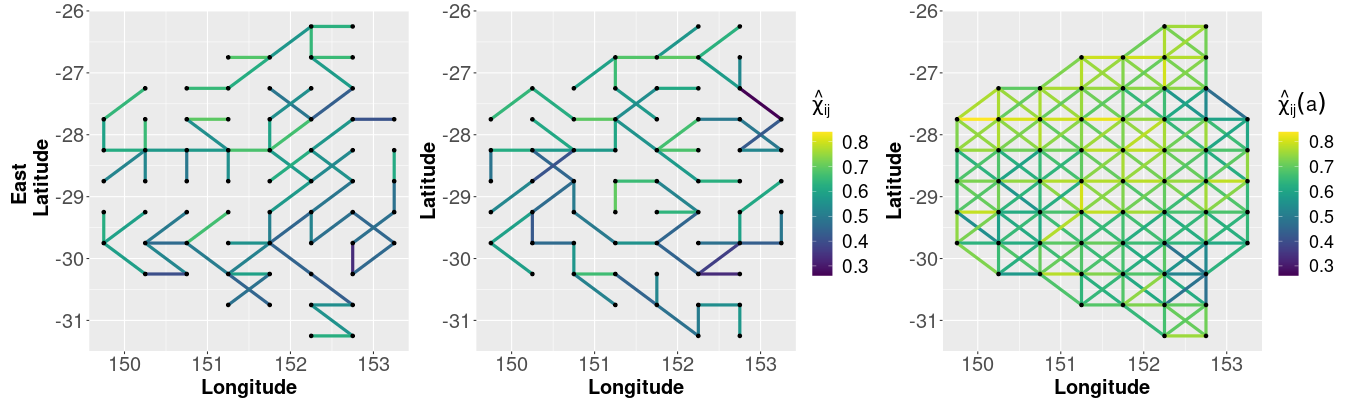}
 \caption{Model-based estimates of $\chi_{ij}$ for the tree-based MPD fitted within two clusters near the western (top) and the eastern (bottom) coastal regions, respectively, for the temporal window 2012--2022. The plots show the estimated tree-based $\chi_{ij}$ for the two trees with the highest (left) and the second highest (middle) estimated mixture probabilities, as well as the optimal $\widehat\chi_{ij}(a_{\rm opt})$ for the tree-mixture MPD model (right).}
 \label{fig:Clus_22_47_win_14}
\vspace{-2mm}
\end{figure}

\begin{figure}[t!]
    \centering
    \begin{subfigure}[t]{0.5\textwidth}
        \centering
        \includegraphics[height=2in]{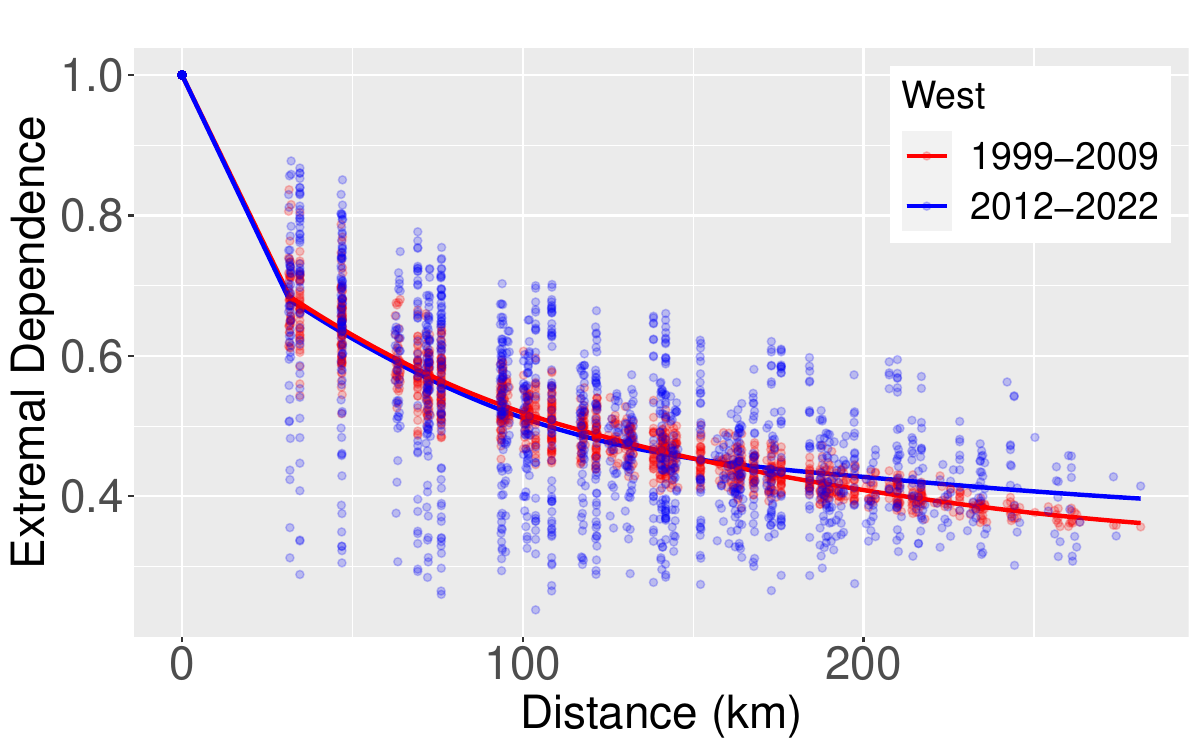}
    \end{subfigure}%
    ~ 
    \begin{subfigure}[t]{0.5\textwidth}
        \centering
        \includegraphics[height=2in]{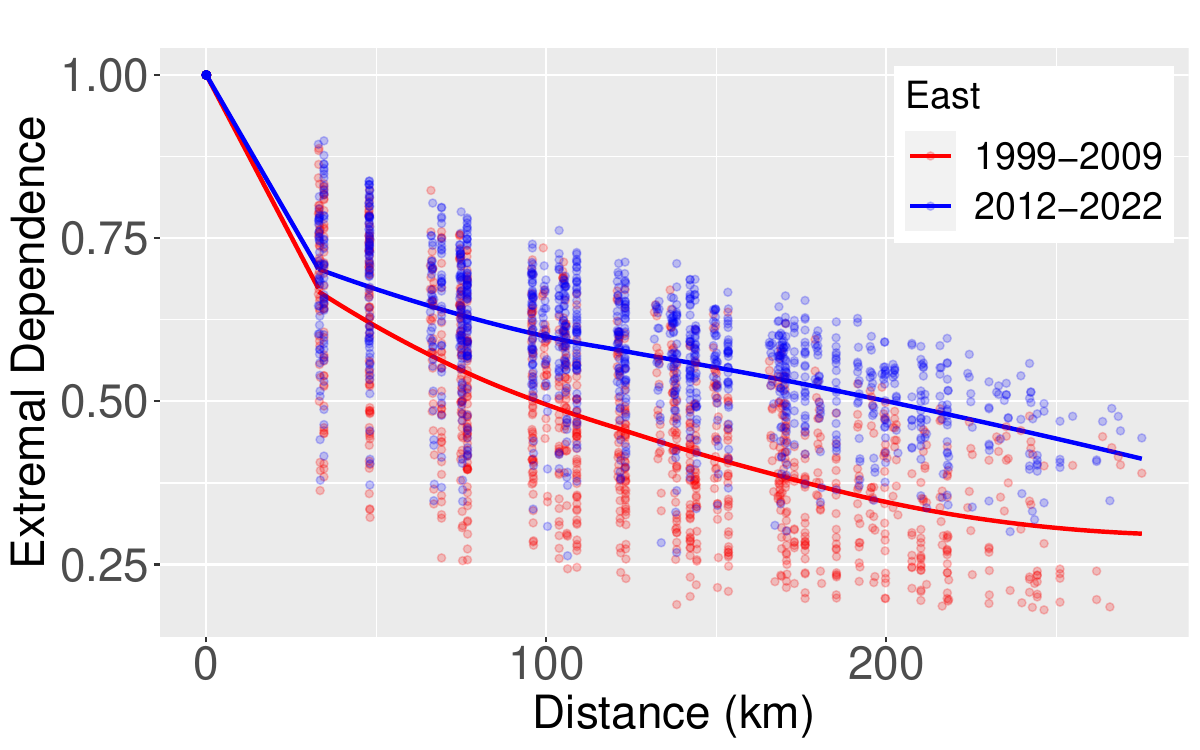}
    \end{subfigure}
    \caption{Smoothed bias-corrected $\chi_{ij}(a)$ estimates as a function of geodesic distance (in km) between sites, based on the proposed tree-mixture MPD model for two spatial clusters along the west (left) and east (right) coasts for the first (red) and last (blue) decade windows.}
    \label{fig:Est_Chi_WE}
 \vspace{-2mm}
\end{figure}

\begin{figure}[t!]
\adjincludegraphics[width = \linewidth, trim = {{.1\width} {.38\width} {.14\width} {.38\width}}, clip]{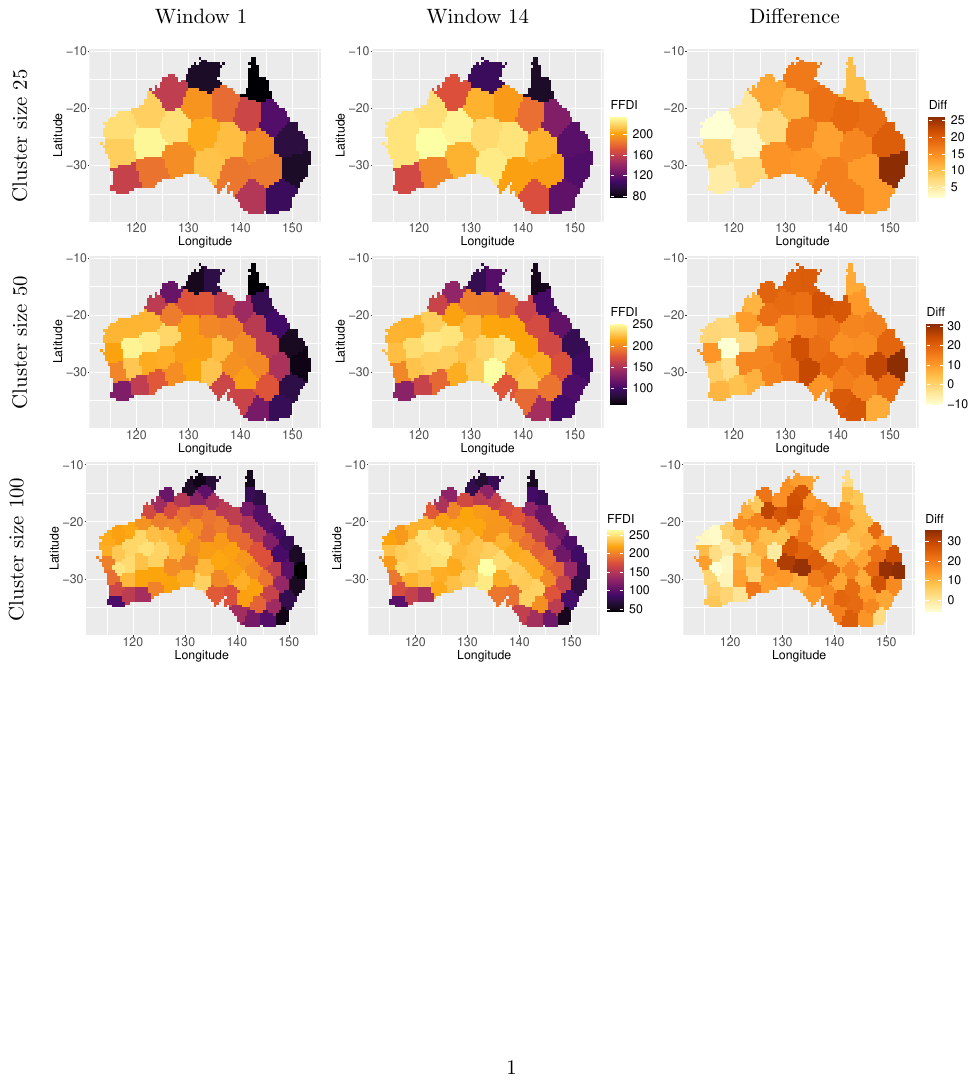}
\caption{Spatial fire risk aggregates for two decade time Windows 1 and 14 (left and middle) and three cluster resolutions. The right panels show the difference in the spatial fire risk aggregates between the two windows.}
\label{fig:Sum_Risk}
 \vspace{-2mm}
\end{figure}

The second step is transform the data to the unit Pareto scale, split the spatial domain into spatial clusters as explained in Section~\ref{sec:DataDescr}, and to fit our proposed MPD tree-mixture model from Section~\ref{subsub:final_model} in each spatio-temporal cluster-window combination. We consider $50$ clusters for most of the analysis, but also consider lower and higher spatial resolutions (with $25$ and $100$ clusters, respectively); see Figure~\ref{fig:Sum_Risk}. The fitted $\chi_{ij}$ measure for the two different clusters (in the southwest and eastern part of the domain) considered in Figure~\ref{fig:FFDI_change_margin_dep} are displayed in Figure~\ref{fig:Clus_22_47_win_14}. The figure shows, for each cluster, the tree-based model estimates of $\chi_{ij}$ for each edge $(i,j)$ of two different trees $\mathcal{T}_m$ with the highest mixture probabilities $p(\mathcal{T}_m)$ (left and middle columns), as well as $\widehat\chi_{ij}(a_{\rm opt})$ for the graph $\mathcal{G}$ based on the optimally bias-corrected tree-mixture MPD model (right column). The estimated dependence structures within these two clusters are clearly non-stationary given that different edges at the same physical distance can have very different $\chi_{ij}$ estimates. 
Figure~\ref{fig:Est_Chi_WE} then illustrates the behavior the estimated $\chi_{ij}$-measure as a function of the geodesic distance (in km) between sites, for the first and last decadal windows and both spatial clusters considered previously. Interestingly, extremal dependence seems to have become stronger over time in the eastern cluster, where major cities are located. This suggests that extreme wildfires on the east coast have become wider in extent, perhaps due to climate change, which has implication for wildfire management and planning. By contrast, we do not see any significant changes for the cluster on the west coast.


Finally, Figure~\ref{fig:Sum_Risk} displays cluster-wise aggregated wildfire risk for the first and last decadal window, computed by averaging samples from the fitted tree-mixture MPD model both over space and independent simulation replicates, after transforming them to the original FFDI scale using the fitted marginal model from Section~\ref{sub:Stage1}. Figure~\ref{fig:Sum_Risk} maps the results for Windows~1 and 14, as well as their difference, and does it for different cluster resolutions (low-resolution with $25$ clusters, mid-resolution with $50$ clusters, and high-resolution with $100$ clusters). We observe generally lower wildfire aggregated risk along the eastern coast of Mainland Australia, but this is also the region where the risk has increased the most. This is concerning particularly because the eastern coast is the most densely populated area of Australia. Our results are helpful to understand the regional impact of climate change on the severity and spatial extent of extreme Australian wildfires, which is key for appropriate risk mitigation and planning. 


\section{Discussions and conclusions}
\label{sec:Conclusions}


In this paper, we introduce a novel approach for analyzing high-dimensional extremes by combining concepts from extreme-value theory and graphical models. Our proposed methodology extends the tree-based extremal graphical models introduced by \cite{engelke2020graphical} to tree-mixtures that may be used for multivariate or spatial extremes. More precisely, the method involves constructing a mixture of tree-based multivariate H\"usler--Reiss-type Pareto distribution models, with appropriately constructed mixture weights defined through a suitable prior over spanning trees, which enables the fast analysis of multivariate or spatial extremes while accounting for nonstationary tail dependence structures. This flexible model effectively captures complex tail dependence structures. 
Our simulation results reported in the Supplementary Material showcase the high flexibility of our modeling approach, and the efficiency of our inference method, as demonstrated by the reasonably good match between the estimated and true extremal dependence profiles in a misspecified setting.



In our real data application, we utilize the newly proposed method to investigate the local spatial tail dependence characteristics of the FFDI fire index in Mainland Australia. To investigate climate change impacts on regional wildfire risk, we adopt a spatio-temporal local approach by analyzing the data separately within spatial clusters and overlapping decadal time windows. Through this approach, we can effectively capture spatio-temporal trends and dependence patterns in FFDI extremes. The results of our analysis provide valuable insights into the severity and spatial extent of extreme wildfire conditions, and how they have changed over time. Our findings highlight that the most densely populated areas of eastern Australia are likely to be most affected by climate change in terms of fire risk. This information is crucial for understanding the potential impacts of climate change on fire-prone regions and can aid in developing targeted mitigation strategies. 

While we here focused on the modeling and risk assessment of wildfires, our methodology applies more generally to other types of natural hazards, such as heavy precipitation or extreme heatwaves. 



\section*{Declarations}
\textbf{Conflicts of interest:} The authors have no conflicts of interest to declare.

\newpage

\baselineskip 14pt
\bibliographystyle{CUP}
\bibliography{references}

\baselineskip 10pt
\newpage




\end{document}